\documentclass[a4paper,11pt]{article}

\usepackage{jcappub} 

\usepackage[T1]{fontenc} 

\usepackage{graphicx}   
\usepackage{amsmath}    
\usepackage{amssymb}    
\usepackage{amsfonts}
\usepackage{xcolor}
\usepackage[normalem]{ulem}

\usepackage{dfpMicros}

\newtheorem{theorem}{Theorem}

\newtheorem{definition}[theorem]{Definition}

\newcommand{\Sp}{\ensuremath{\mathbb{S}}}
\newcommand{\bE}{\ensuremath{\mathbb{E}}}

\newcommand{\bn}{\ensuremath{\mathbf{n}}}
\newcommand{\bx}{\ensuremath{\mathbf{x}}}

\newcommand{\bp}{\ensuremath{\mathbf{p}}}
\newcommand{\bz}{\ensuremath{\mathbf{z}}}
\newcommand{\intd}[1]{{\rm d}\sigma(#1)}
\newcommand{\expect}[1]{\mathbb{E}\left[#1\right]} 
\newcommand{\Exp}[1]{\mathbb{E}\left[ #1 \right]}
\definecolor{dgreen}{rgb}{0.0,0.5,0.0}
\definecolor{borange}{rgb}{0.8,0.33,0.0}

\title{\boldmath A New Probe of Gaussianity and Isotropy applied to the CMB Maps}

\author[a]{J. Hamann,}
\author[b,1]{Q. T. Le Gia,\note{Corresponding author.} }
\author[b]{I. H. Sloan,}
\author[b]{Y. G. Wang}
\author[b]{and \\R. S. Womersley}

\affiliation[a]{School of Physics, The University of New South Wales, \\
Sydney, NSW 2052, Australia}
\affiliation[b]{School of Mathematics and Statistics, The University of New South Wales,\\Sydney, NSW 2052, Australia}

\emailAdd{jan.hamann@unsw.edu.au}
\emailAdd{qlegia@unsw.edu.au}
\emailAdd{i.sloan@unsw.edu.au}
\emailAdd{yuguang.wang@unsw.edu.au}
\emailAdd{r.womersley@unsw.edu.au}

\abstract{We introduce a new mathematical tool (a direction-dependent probe) to
analyse the randomness of purported isotropic Gaussian random fields
on the {sphere}. We apply the probe to assess the full-sky cosmic
microwave background (CMB) temperature maps produced by the {\it Planck} collaboration (PR2 2015 and PR3 2018), with special attention to the inpainted maps. To study the randomness of the fields represented by each map we use the autocorrelation of the sequence of probe coefficients (which are just the full-sky Fourier coefficients $a_{\ell,0}$ if the $z$ axis is taken in the probe direction).  If the field is {isotropic and
Gaussian} then the probe coefficients for a given direction should be
realisations of uncorrelated scalar Gaussian random variables. We introduce a particular function on the sphere (called the \emph{AC discrepancy}) that accentuates the departure from Gaussianity and isotropy. We find that for some of the maps, there are many directions for which the departures are significant, especially near the galactic plane.  We also study the effect of varying the highest multipole used to calculate the AC discrepancy from the initial value of $1500$ to $2500$.
In the case of Commander 2015, the AC discrepancy now exhibits
antipodal ``blobs'' well away from the galactic plane.  Finally, we look briefly at the non-inpainted Planck maps, for which the computed AC discrepancy maps have a very different character, with features that are global rather than local.  For the particular case of the non-inpainted 2018 \texttt{SEVEM} map (which has visible equatorial pollution), we model with partial success the observed behaviour by an isotropic Gaussian random field added to a non-random needlet-like structure located near the galactic centre.}

\begin{document}
\maketitle
\flushbottom

\section{Introduction}
\label{sec:intro}
In this paper, we introduce a mathematical tool (henceforth referred to as the ``probe'') for analysing the randomness of purported realisations of an isotropic Gaussian random field; and apply it to full-sky cosmic microwave background (CMB) maps produced by the {\it Planck} consortium \citep{Planck2015IX, Planck2018iv}.
This probe should be regarded as complementary to other methods of searching for non-Gaussianity~\citep{Akrami:2019izv} or deviations from statistical isotropy of the CMB~\citep{Akrami:2019bkn}.
It relies on the Fourier coefficients, even those of high degree,
being easily computed. (The computation of approximate Fourier coefficients in the case of the {\it Planck} 
full-sky maps is straightforward even up to multipole degrees $\ell > 2500$, because they are given at the
\texttt{HEALPix}\footnote{\url{http://healpix.sf.net}} points
\citep{Gorski_etal2005}, which are designed to make the harmonic
analysis efficient.)

Assuming that the origin of the primordial density fluctuations lies in a phase of generic slow-roll inflation~\citep{Mukhanov:991646}, the {\it primary} CMB anisotropies (i.e., the temperature fluctuations on the surface of the last scattering) can be very well described as a realisation of a statistically isotropic Gaussian random field on the $2$-dimensional sphere.  Note, however, that the observed CMB is a superposition of
the primary anisotropies and {\it secondary} anisotropies~\citep{Aghanim:2007bt} generated due to the propagation
through an anisotropic medium, and that the secondary anisotropies are
expected to deviate from Gaussianity, most notably due to weak
gravitational lensing~\citep{Lewis:2006fu} and the Sunyaev-Zel'Dovich
effect~\citep{Sunyaev:1980nv}.

\begin{figure}
\centering
\includegraphics[trim = 0mm 0mm 0mm 0mm, width=0.75\columnwidth]{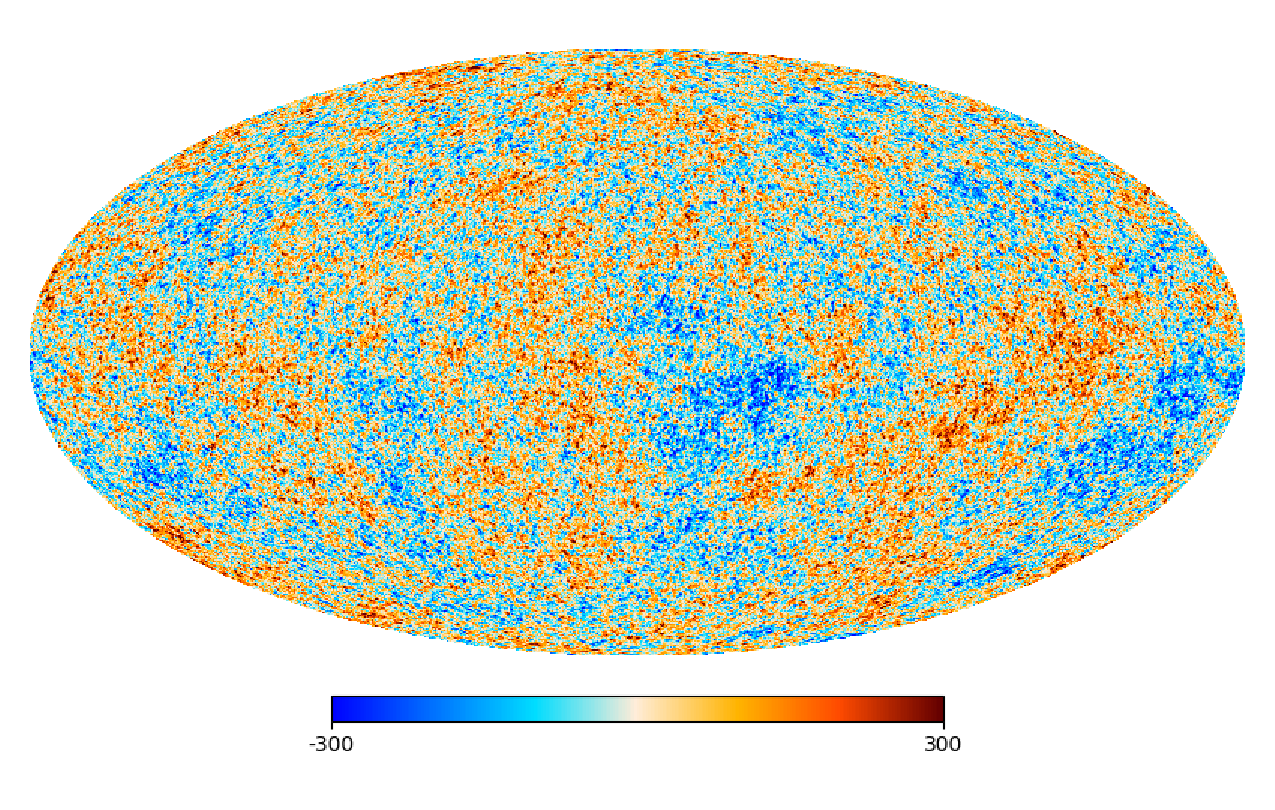}\vspace{-5mm}
\caption{Inpainted CMB map, \texttt{Commander} 2015, $N_{\textrm{Side}} = 2048$}
\label{fig:cmb_commander}
\end{figure}

\begin{figure}
\centering
\includegraphics[trim = 0mm 0mm 0mm 0mm, width=0.75\columnwidth]{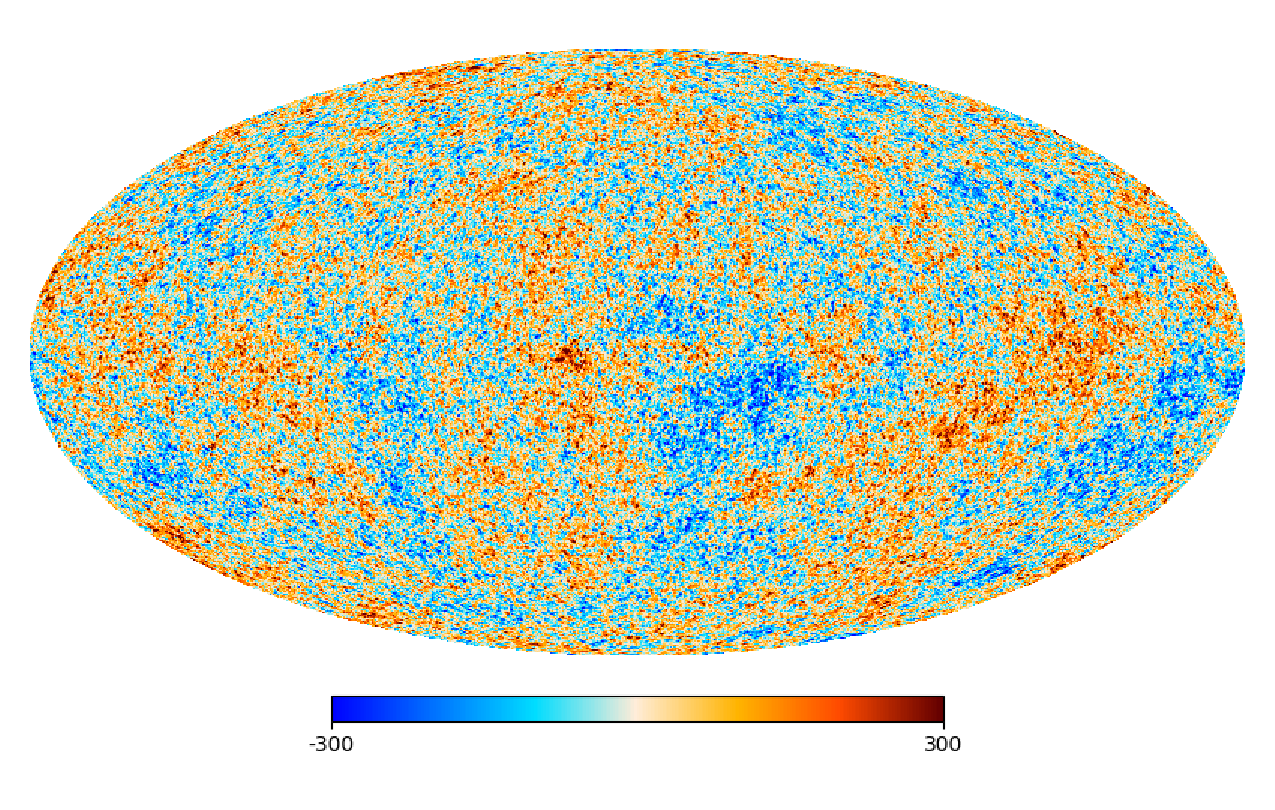}\vspace{-5mm}
\caption{Inpainted CMB map, \texttt{SEVEM} 2018, $N_{\textrm{Side}} = 2048$}
\label{fig:cmb_sevem2018}
\end{figure}

The tool we shall introduce is a highly directional axially symmetric spherical harmonic of arbitrary degree $\ell$, with axis in the direction of an arbitrary unit vector $\bp$.  When convolved with the field it provides sensitive information about the extent to which the field, as represented by the map, is truly Gaussian and isotropic.  We shall concentrate initially on the inpainted versions of the CMB maps because the non-inpainted versions often display visible pollution from foreground effects in the masked region near the galactic equator.  The inpainted versions, in contrast, all appear good to the eye. In Figure~\ref{fig:cmb_commander}\footnote{The data sets for the CMB
maps used in this paper were downloaded from the {\it Planck} Legacy Archive at \url{https://pla.esac.esa.int/\#maps}.} we show the \texttt{Commander} 2015 temperature map in a Mollweide projection. In Figure~\ref{fig:cmb_sevem2018} we show the inpainted version of the 2018 \texttt{SEVEM} map. The other 2018 inpainted maps are similar.

A real-valued scalar field $T$ on the unit sphere $\mathbb{S}^2$ in
$\mathbb{R}^3$ can be expanded in terms of its Fourier series
\begin{equation}\label{eq:Texp1}
T(\bx) = \sum_{\ell=0}^\infty\sum_{m=-\ell}^\ell a_{\ell,m}
Y_{\ell,m}(\theta,\phi),
\end{equation}
where $Y_{\ell,m}$ are the usual orthonormal family of complex
spherical harmonics 
of degree $\ell$, $\theta$ is the polar angle of $\bx$, and $\phi$
the azimuthal angle. (The polar angles are related to the galactic
coordinates $(l, b)$ by $\theta = \pi/2 - l$ and $\phi = b$.) In terms of
$T$, the Fourier coefficients $a_{\ell,m}$ are given, for $m =
-\ell,\ldots,\ell$ and $\ell = 0,1, \ldots$, by
\begin{equation}\label{eq:a_lm}
a_{\ell,m}= \int_{\mathbb{S}^2} \overline{Y_{\ell,m}(\theta,\phi)} T(\bx) \intd{\bx}.
\end{equation}
Here $\intd{\bx}$ is the surface measure for the $2$-dimensional unit sphere $\mathbb{S}^2$.

For a strongly isotropic Gaussian random field $T$ (for the
definition see Section~\ref{sec:math.background}), the Fourier
coefficients $\{a_{\ell,m}: m = -\ell,\dots,\ell, \ell = 1,2,\dots \}$ are
uncorrelated complex-valued Gaussian random variables which satisfy
\begin{equation}\label{mean var alm}
\expect{a_{\ell,m}}=0,\quad \expect{a_{\ell,m}\overline{a_{\ell',m'}}} =C_{\ell}\delta_{\ell,\ell'}\delta_{m,m'},
\end{equation}
where $\delta_{j,k}$ is the Kronecker symbol and the positive constant $C_{\ell}$ depends on $\ell$ only. The sequence $C_{\ell}$ is known as the \emph{angular power spectrum} of the field. As is usual in the CMB context, we require for the monopole and dipole $C_0 = C_1= 0$. The set of the $a_{\ell,m}$ with $\ell\ge 2$ and $m\ge 0$ are in the Gaussian case not just uncorrelated but are independent Gaussian random variables. It is easily seen that the Gaussian field $T$ has mean zero,
\[
\mathbb{E}[T(\bx)] = 0, \quad \bx \in \mathbb{S}^2,
\]
and covariance
\begin{equation}\label{eq:cov}
\mathbb{E}[T(\bx)T(\bz)] = \sum_{\ell=
2}^\infty C_\ell \frac{2\ell+1}{4\pi} P_\ell(\bx\cdot\bz), \quad \bx, \bz
\in \mathbb{S}^2,
\end{equation}
where $P_\ell$ is the Legendre polynomial of degree $\ell$, and we used the addition theorem \citep{Muller1966}.

\begin{equation}\label{addition}
P_\ell(\bx\cdot \bz) =
\frac{4\pi}{2\ell+1}\sum_{m=-\ell}^\ell Y_{\ell, m}(\bx)\overline{Y_{\ell, m}(\bz)},
\quad \bx, \bz \in \mathbb{S}^2.
\end{equation}

For the moment we concentrate on the coefficients $a_{\ell,0}$ (i.e. we take $m=0$), but we consider simultaneously many different $\ell$ values, recalling  that the $a_{\ell,0}$  are to a good approximation supposed to be instances of \emph{independent} random variables. It is useful to consider exactly what the coefficients $a_{\ell,0}$ tell us about a field $T$. Remembering that
\begin{equation}\label{eq:Yl0}
        Y_{\ell,0}(\theta, \phi)= \sqrt{\frac{2\ell+1}{4\pi}} P_\ell(\cos\theta) =\sqrt{\frac{2\ell+1}{4\pi}} P_\ell(\bx\cdot\bn),
\end{equation}
where $\bn$ is the unit vector in the direction of the positive $z$ axis, it follows from \eqref{eq:a_lm} that $a_{\ell,0}$ is a real number given by
\begin{equation}\label{eq:al0}
    a_{\ell,0} = \int_{\sph{2}}\sqrt{\frac{2\ell+1}{4\pi}}P_\ell(\bx\cdot\bn)T(\bx)\intd{\bx}.
\end{equation}
Thus $a_{\ell,0}$  is the convolution of $T$ with a spherical harmonic of degree $\ell$ that is axially symmetric about the $z$-axis.
\begin{figure}[ht]
\centering
\includegraphics[width=0.75\columnwidth]{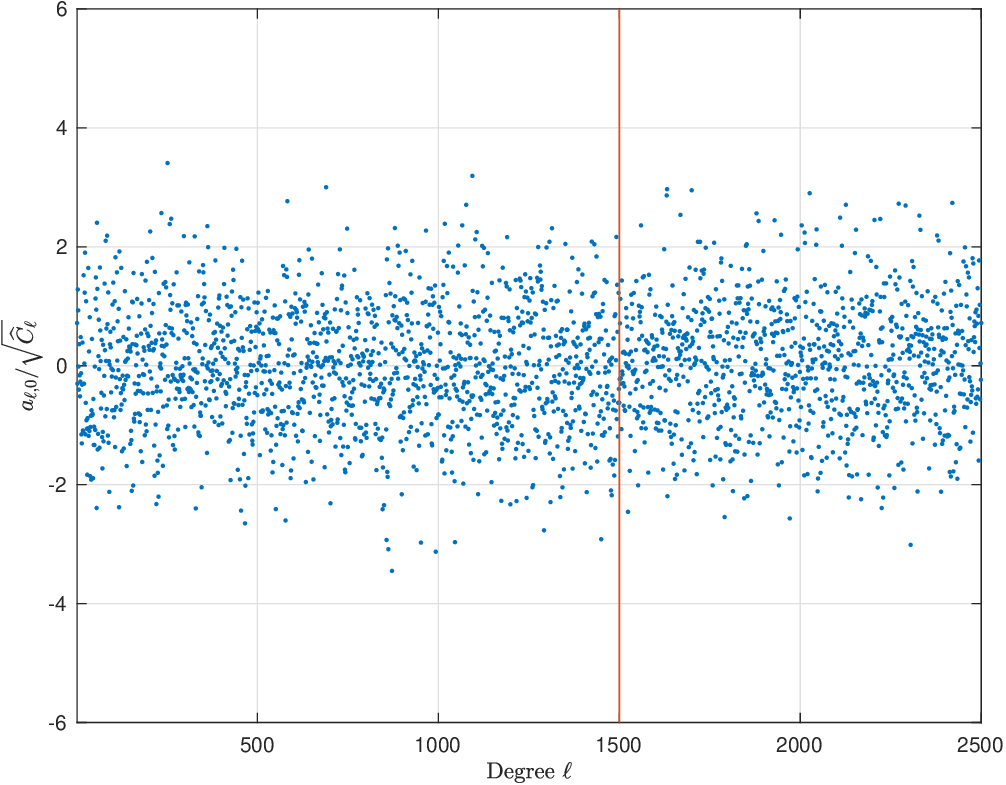}\\[2mm]
\caption{Coefficients $a_{\ell,0}/\sqrt{\widehat{C}_\ell}$ for inpainted \texttt{Commander} 2015}
\label{fig:comm_al0}
\end{figure}
\begin{figure}[ht]
\centering
\includegraphics[width=0.75\columnwidth]{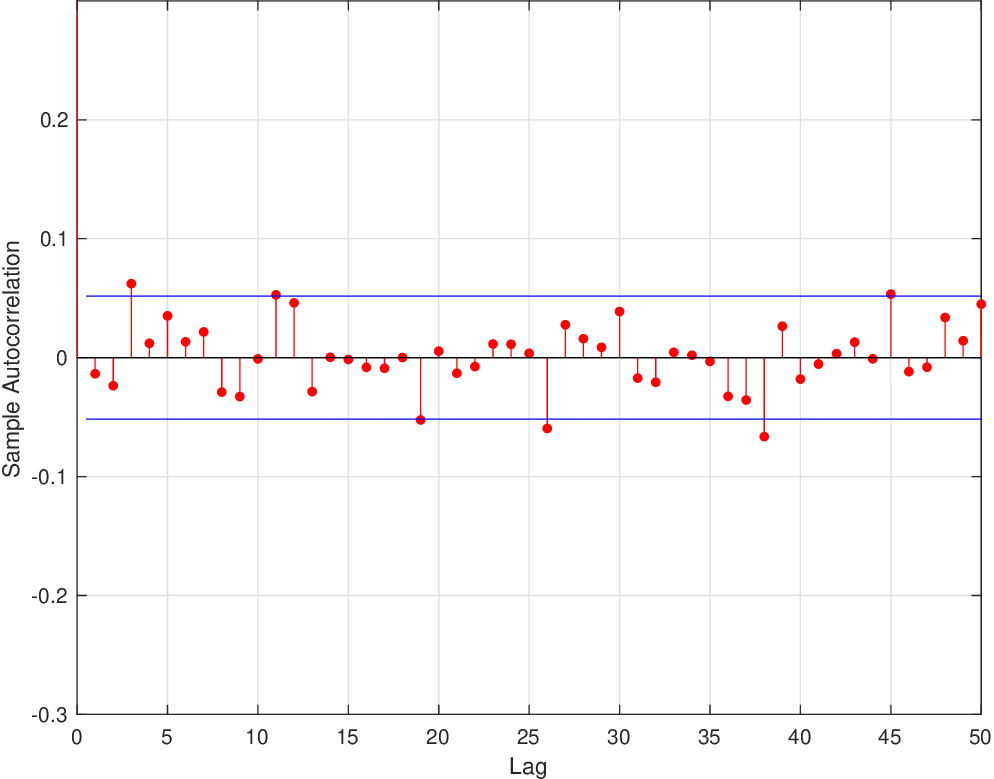}\vspace{2mm}
\caption{Autocorrelations of $a_{\ell,0}/\sqrt{\widehat{C}_\ell}$ for inpainted \texttt{Commander} 2015}
\label{fig:comm15_autocorr}
\end{figure}

In Figure \ref{fig:comm_al0} we show for the \texttt{Commander} 2015 map the (real) numbers $a_{\ell,0}$ divided by $\sqrt{\widehat{C}_\ell}$, for values of $\ell$ from $2$ up to $2,500$. (The vertical line at $\ell =
1500$ shows the upper limit of the $\ell$ values used in most of the
paper, until Section\ref{sec:compareL}.) Here the squared normalising factor
$\widehat{C}_\ell$ is an empirical estimate of $C_\ell$, defined in
terms of the computed $a_{\ell,m}$ values by
\begin{equation}\label{eq:empiricalC}
	\widehat{C}_\ell:=\frac{1}{2\ell+1} \sum_{m=-\ell}^\ell |a_{\ell,m}|^2,\quad\ell\ge 0.
\end{equation}
If the Gaussianity assumption holds then the scaled coefficients
$a_{\ell,0}/\sqrt{\widehat{C}_\ell}$ should be instances of independent
mean-zero Gaussian random variables with variance close to $1$. To the eye this appears satisfactorily to be the situation for the $a_{\ell,0}$ from
\texttt{Commander} 2015 in Figure~\ref{fig:comm_al0}.

The one-sample Kolmogorov-Smirnov test gives a quantitative measure of the null hypothesis for studying the randomness of a sequence, which in this case is that the scaled coefficients $a_{\ell,0} /\sqrt{\widehat{C}_\ell}$
are independent samples from a standard normal distribution.
It gives a $p$-value of $0.80$ for the \texttt{Commander} 2015 data in Figure~\ref{fig:comm_al0}, indicating that the null hypothesis cannot be rejected for this data set.

To find if a sequence of similarly scaled real numbers is correlated or
uncorrelated, a standard device in the world of time series is to compute
the \emph{autocorrelation} for lags of $1, 2, \ldots$. (For a definition
of autocorrelation and full information on the computation, see
Section~\ref{sec:math.background}.) In Figure \ref{fig:comm15_autocorr} we compute the autocorrelations for the data in Figure \ref{fig:comm_al0} for all lags from $1$ up to $50$, using all the multipoles
from $\ell=2$ to $1500 =:L$.\footnote{{Our choice of maximum multipole $L$ for the AC discrepancy analysis is informed by the requirement that we stay well within the range where \textit{Planck} maps are dominated by the CMB.  Beyond $\ell \gtrsim 1600$, the \textit{Planck} maps start becoming dominated by noise~\cite{Akrami:2018vks}.  Owing to \textit{Planck}'s scanning pattern, the noise represents an anisotropic contribution to the maps, which would manifest itself as a spurious signal in the AC discrepancy.}} (The autocorrelation for lag $0$ by definition always has the value $1$.) The two horizontal blue lines in
Figure~\ref{fig:comm15_autocorr} are at $\pm t_L$, where
\begin{equation}\label{eq:threshold}
t_L =2/\sqrt{L-1} = 0.0517
\end{equation}
is the $95.45\%$ confidence interval if the input data
consist of iid standard normal random variables.
If the hypothesis holds that the input data are iid normal random variables, the autocorrelations themselves are uncorrelated, and almost all of the autocorrelations lie between the blue lines.
That indeed seems to be qualitatively the case for the data from \texttt{Commander} 2015 in Figure~\ref{fig:comm15_autocorr}.

The coefficients $a_{\ell,0}$ are seen in \eqref{eq:al0} to be associated
with the direction $\bn$ of the positive $z$-axis. But the $z$-axis
is not the only interesting direction. We now define the probe, which will
allow us to test the field for any direction $\bp$, with $\bp$ an
arbitrary unit vector. The probe is a real-valued mathematical function on
the sphere of the simple form
\begin{equation}\label{eq:probedef}
\mathcal{P}_{\ell,\bp}(\bx):=
\sqrt{\frac{2\ell+1}{4\pi}}P_\ell(\bx\cdot\bp), \quad \ell = 0,1, \ldots,
\end{equation}
which is a spherical harmonic of degree $\ell$, rotationally symmetric about an axis in the direction of $\bp$. The \emph{probe coefficient for the direction} $\bp$ is the inner product of $\mathcal{P}_{\ell,\bp}$ with the given real scalar
field $T(\bx)$,
\begin{equation}\label{eq:Tptobe} T_{\ell,\bp} :=
\int_{\mathbb{S}^2} \mathcal{P}_{\ell,\bp}(\bx) T(\bx)\intd{\bx}, \quad
\ell = 0,1, \ldots.
\end{equation}
In the special case when $\bp = \bn = (0,0,1)$ we have
\[
\mathcal{P}_{\ell,\bp}(\bx) =\sqrt{\frac{2\ell+1}{4\pi}}P_\ell(\cos\theta)
= Y_{\ell,0}(\theta, \phi) \quad\mbox{and}\quad
T_{\ell,\bp} = a_{\ell,0}.
\]
To make the same point differently, if the $z$-axis is chosen in the direction of $\bp$ then the probe $\mathcal{P}_{\ell,\bp}$ is just the spherical harmonic of degree $\ell$ with $m=0$, and the probe coefficient $T_{\ell,\bp}$ is just $a_{\ell,0}$.

The autocorrelation for probe direction $\bp$ is intimately connected to that at its antipode $-\bp$: for we shall see in the next section that the probe coefficient has the symmetry property
\begin{equation}\label{eq:symmTlp}
T_{\ell,-\bp}=(-1)^\ell T_{\ell,\bp}.
\end{equation}
The scaled probe coefficients are
\begin{equation}\label{eq:wtdT}
        \widetilde{T}_{\ell,\bp}:=T_{\ell,\bp}/\sqrt{\widehat{C}_{\ell}},
        \quad \ell=2,\dots,L.
\end{equation}
\clearpage

\begin{figure}
\vskip 2mm
\centering
\includegraphics[width=0.75\columnwidth]{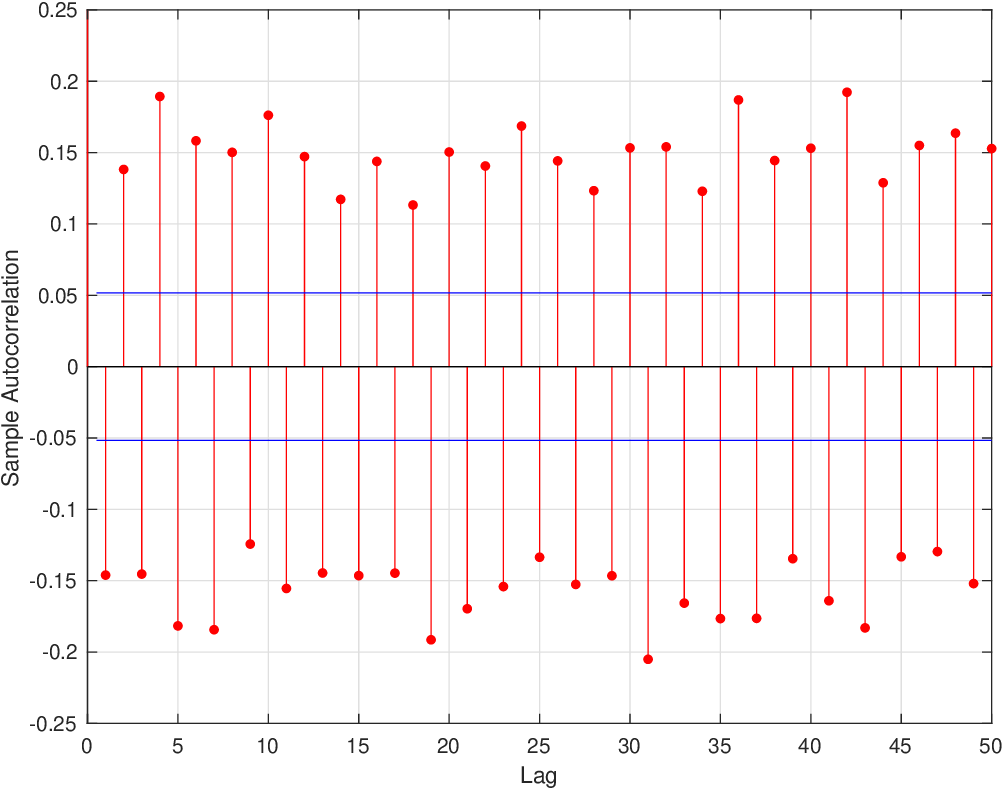}\\[1mm]
\caption{Autocorrelations of $\widetilde{T}_{\ell,\bp}$, $\bp$ at $(l,b) = (353.54, 1.79)$,
which gives the maximum value $1.077$ of the AC discrepancy with $L = 1500$ for the \texttt{Commander} 2015 map.}
\label{fig:comm15_ac_maxacd}
\end{figure}
\vspace{108mm}
\begin{figure}
\vskip 2mm
\centering
\includegraphics[width=0.75\columnwidth]{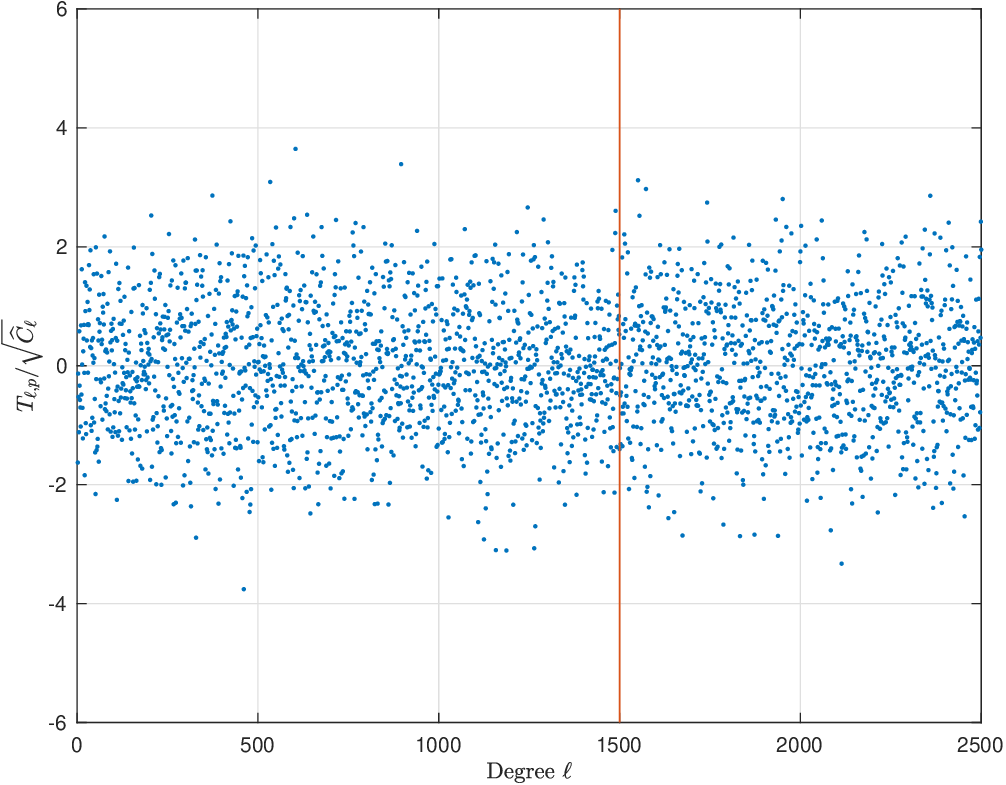}\\[1mm]
\caption{Scaled probe coefficients $\widetilde{T}_{\ell,\bp}$, $\bp$ at $(l,b) = (353.54, 1.79)$, 
\texttt{Commander} 2015}
\label{fig:comm15_coeff_maxacd_anti}
\end{figure}

An example of a direction $\bp$ for which the probe for \texttt{Commander} 2015 reveals a significant departure from randomness is, in galactic coordinates, $(l,b) = (353.54, 1.79)$.
For this direction the autocorrelations of the scaled probe coefficients $\widetilde{T}_{\ell,\bp}$
given in Figure \ref{fig:comm15_ac_maxacd} show strong correlation even up to a lag of $50$.
For completeness, the underlying scaled probe coefficients are shown in
Figure~\ref{fig:comm15_coeff_maxacd_anti}. We explain below how this direction was discovered.

To allow us to explore the {\it Planck} temperature maps for all
directions $\bp$, so seeking regions where the autocorrelations of scaled
values of $T_{\ell,\bp}$ depart significantly from the expected behaviour,
we compute for each such temperature map the \emph{autocorrelation
discrepancy} (defined in Section~\ref{sec:math.background}) of the numbers $T_{\ell,\bp}$, with $\bp$ varying over the sphere. It is a measure of the extreme departure of the autocorrelation from that of a Gaussian distribution, designed to allow us to detect regions (or the antipodes of such regions) in which the autocorrelation departs most strongly from the Gaussian assumption. In the case of \texttt{Commander} 2015 the AC discrepancy reaches a maximum value of $1.077$ over $12,582,912$ HEALPix points (with $N_{\rm Side} = 1024$) on the sphere. The corresponding autocorrelations are shown in Figure~\ref{fig:comm15_ac_maxacd}.

\begin{figure}
\vspace{-12mm}
\centering
\includegraphics[trim = 0mm 0mm 0mm 0mm, width=0.75\columnwidth]{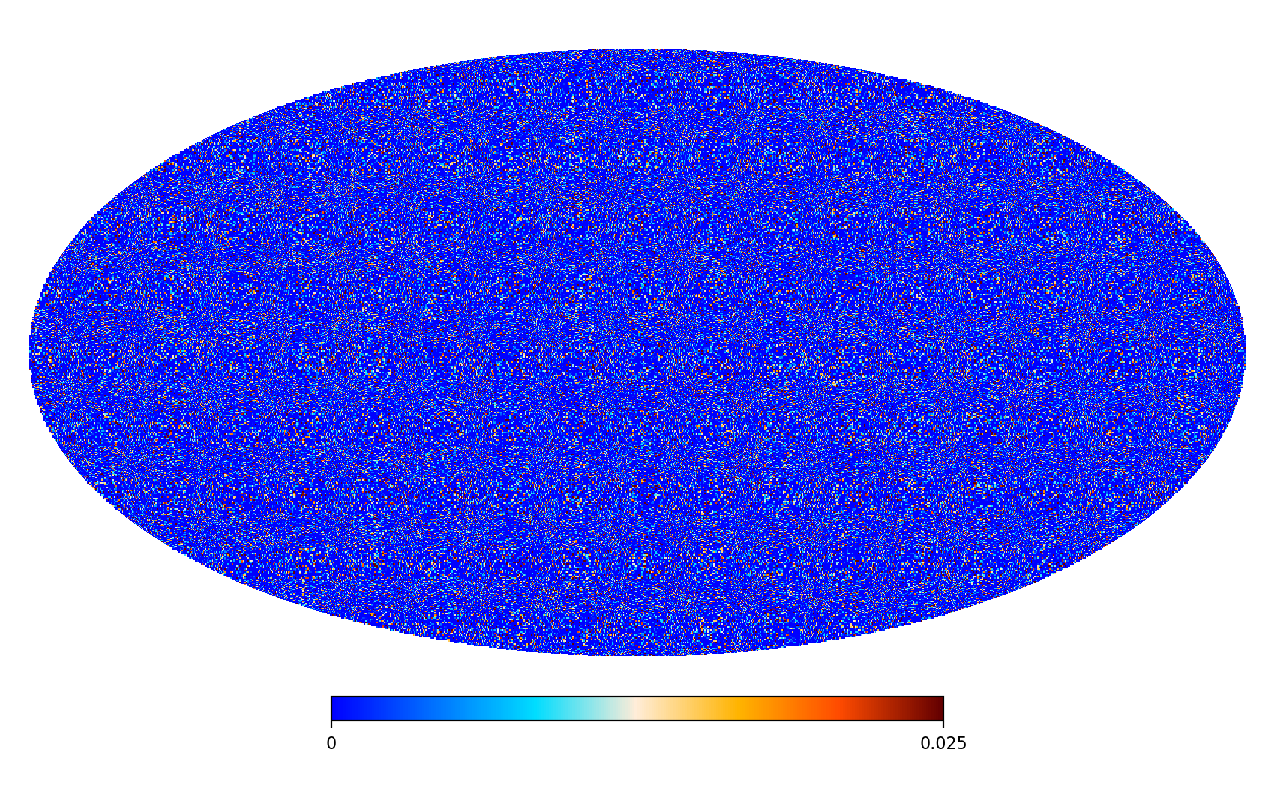}\\[-4mm]
\caption{AC discrepancy map, \texttt{Commander} 2015, $N_{\textrm{Side}} = 1024$,
$k_{\max} = 10$, $L = 1500$, maximum $1.077$ at $(l,b) = (353.54, 1.79)$}
\label{fig:acdmap_comm15}\vspace{3mm}
\includegraphics[trim = 0mm 0mm 0mm 0mm, width=0.75\columnwidth]{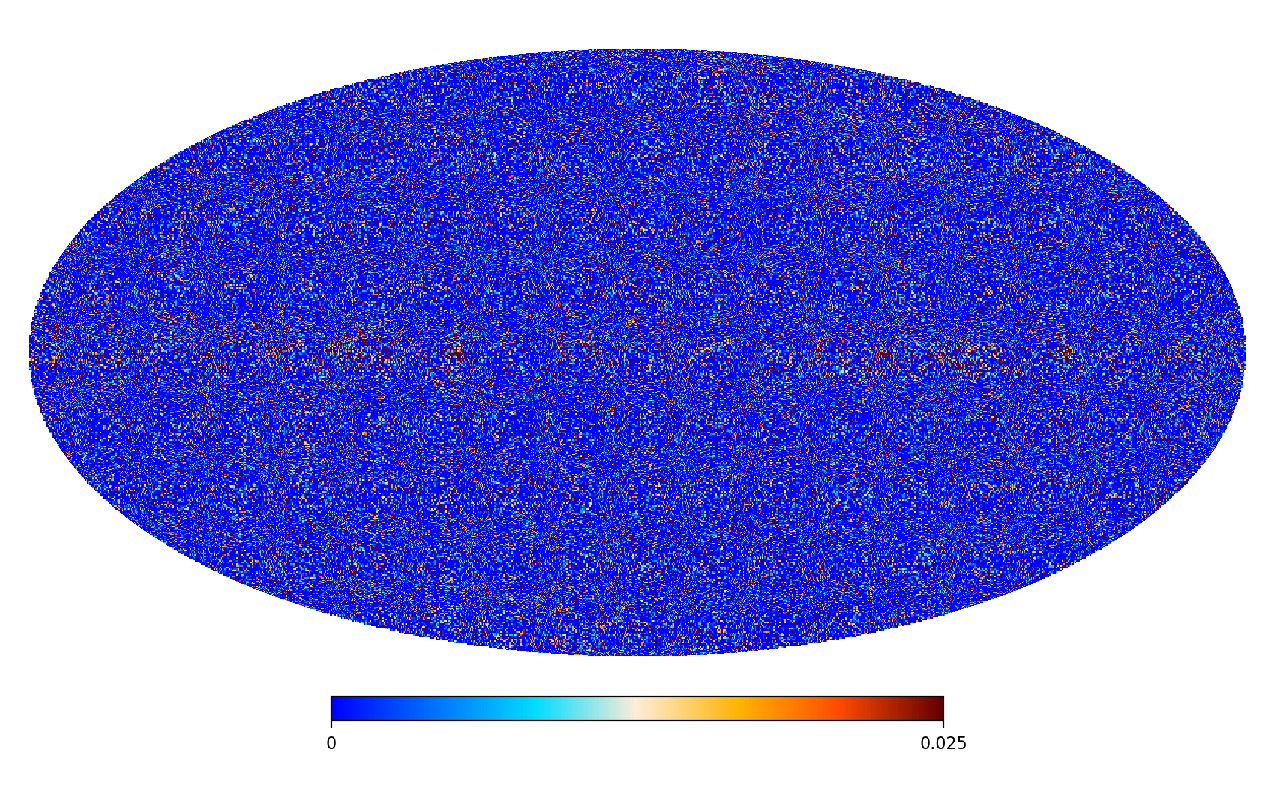}\\[-4mm]
\caption{AC discrepancy map, inpainted \texttt{Commander} 2018, $N_{\textrm{Side}} = 1024$,
$k_{\max} = 10$, $L = 1500$, maximum $2.631$ at $(l,b) = (12.57, 0.11)$}
\label{fig:acdmap_comm18}\vspace{3mm}
\includegraphics[trim = 0mm 0mm 0mm 0mm, width=0.75\columnwidth]{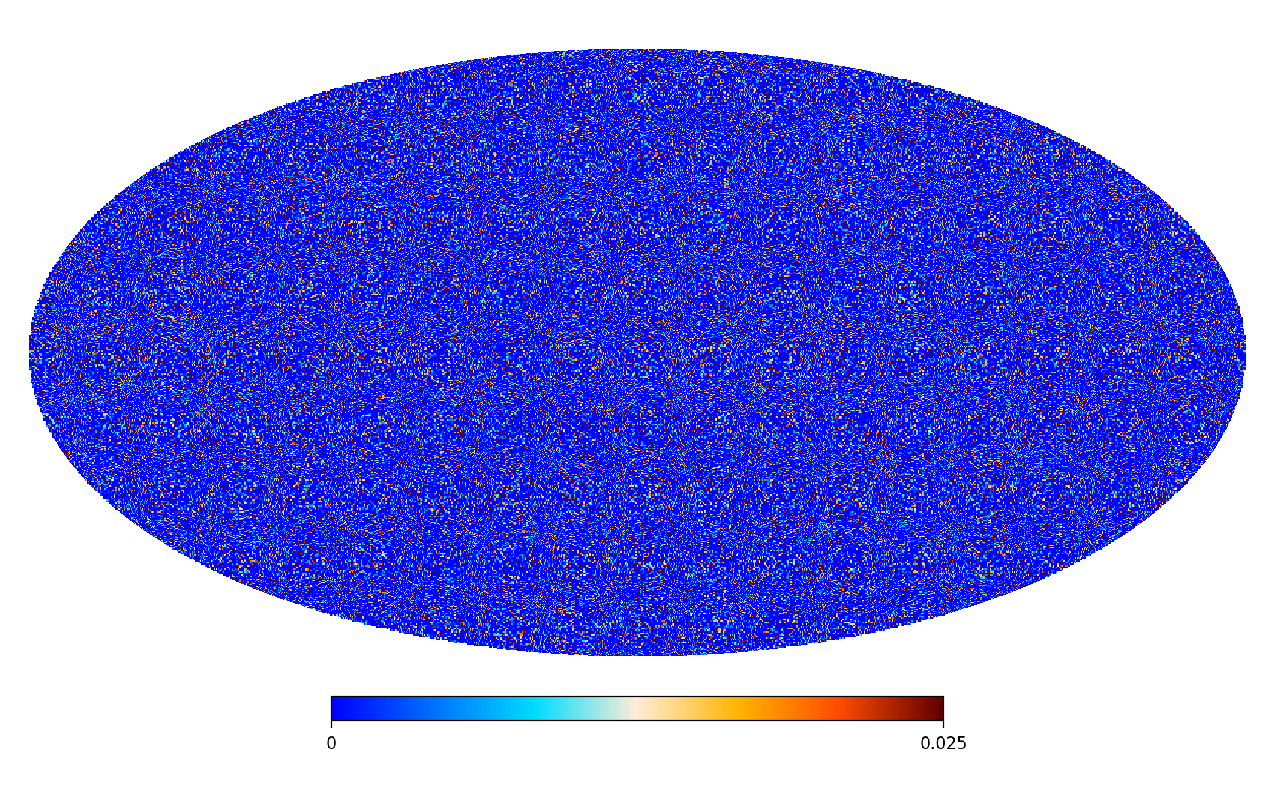}\\[-4mm]
\caption{AC discrepancy map, inpainted \texttt{NILC} 2018, $N_{\textrm{Side}} = 1024$,
$k_{\max} = 10$, $L = 1500$, maximum $0.134$ at $(l,b) = (61.17, -30.73)$}
\label{fig:acdmap_nilc18}
\end{figure}

\begin{figure}
\vspace{-8mm}
\centering
\includegraphics[trim = 0mm 0mm 0mm 0mm, width=0.75\columnwidth]{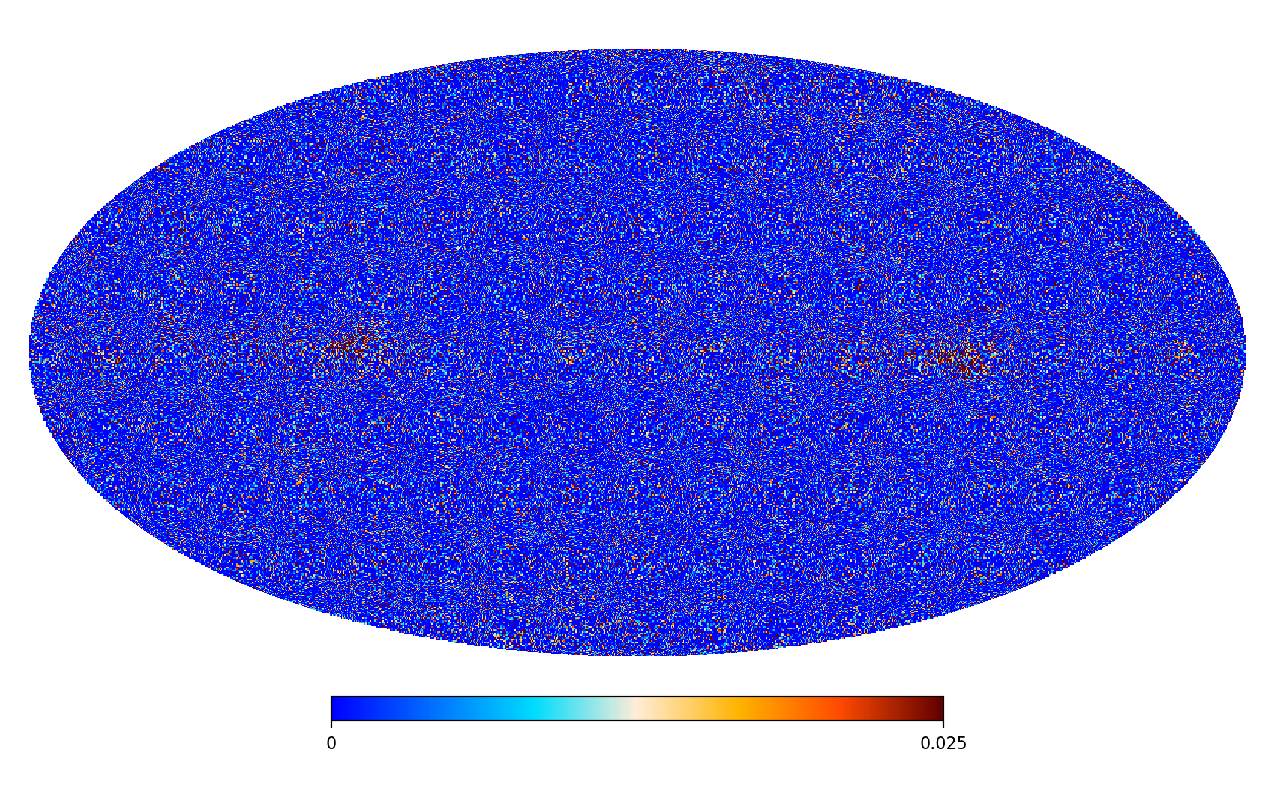}\\[-4mm]
\caption{AC discrepancy map, inpainted \texttt{SEVEM} 2018, $N_{\textrm{Side}} = 1024$,
$k_{\max} = 10$, $L = 1500$, maximum $0.270$ at $(l,b) = (261.25, -2.99)$}
\label{fig:acdmap_sevem18}
\end{figure}

\begin{figure}
\vspace{-2mm}
\centering
\includegraphics[trim = 0mm 0mm 0mm 0mm, width=0.75\columnwidth]{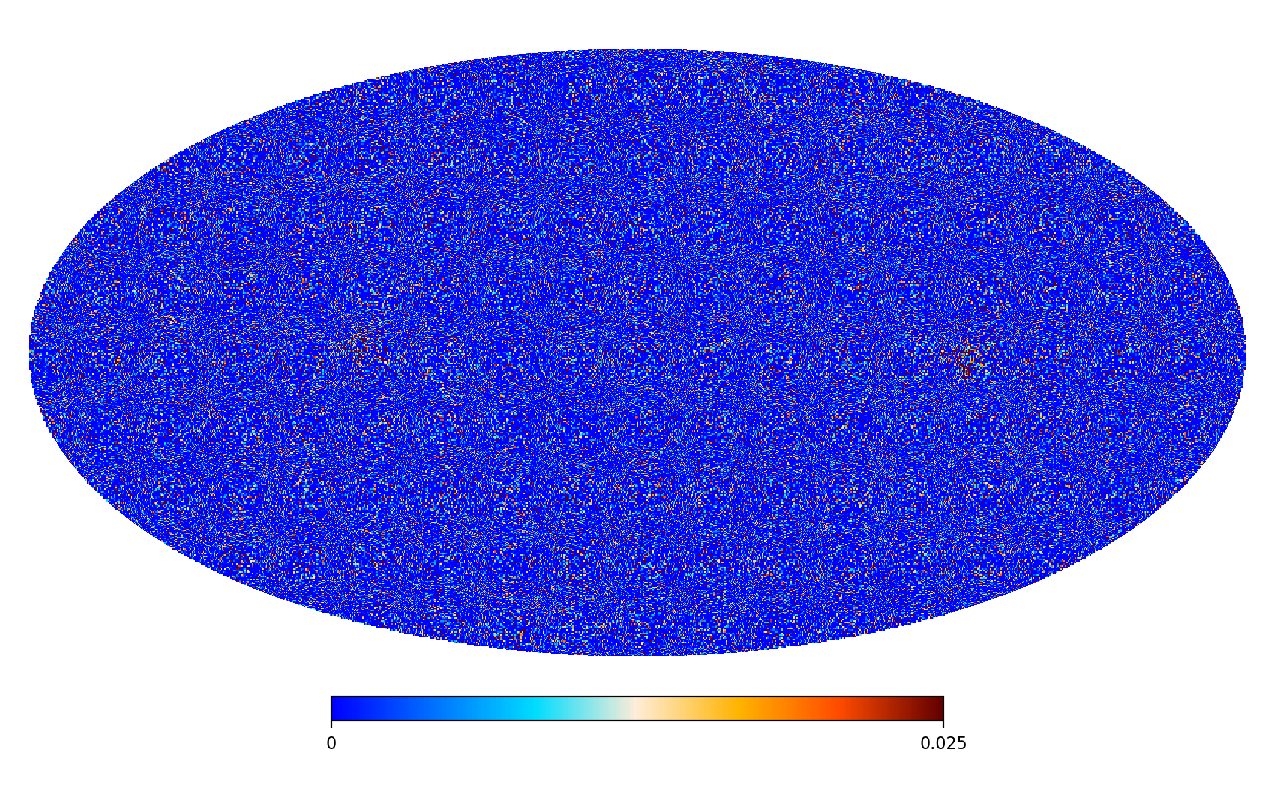}\\[-4mm]
\caption{AC discrepancy map, inpainted \texttt{SMICA} 2018, $N_{\textrm{Side}} = 1024$,
$k_{\max} = 10$, $L = 1500$, maximum $0.397$ at $(l,b) = (261.34, -2.99)$}
\label{fig:acdmap_smica18}
\end{figure}

In Figures \ref{fig:acdmap_comm15}, \ref{fig:acdmap_comm18},
\ref{fig:acdmap_nilc18}, \ref{fig:acdmap_sevem18} and
\ref{fig:acdmap_smica18} we show the AC discrepancy maps for
\texttt{Commander} 2015 and all four of the inpainted 2018 maps, all with
the same colour map. We observe that the AC discrepancy maps for \texttt{NILC} and \texttt{Commander} 2015 do not seem to display any obvious directional preference whereas \texttt{Commander} 2018, \texttt{SEVEM} and \texttt{SMICA} have visible
pollution in the inpainted region.

A general observation about all of the AC discrepancy maps in
Figures~\ref{fig:acdmap_comm15}--\ref{fig:acdmap_smica18} is that the
larger values (that is, the larger departures from the background blue)
appear to be randomly scattered, with no continuity between one
pixel and the next. (We shall see later, in Section~\ref{sec:non-inpainted}, that the corresponding AC discrepancies of the non-inpainted maps have a completely different character --- they are larger, wilder, but essentially continuous.) That being the case, it follows that (contrary to our expectation) there is no special significance attached to the precise point at which the AC discrepancy happens to reach its maximum value --- the large value may be just a consequence of random fluctuation. On the other hand, the visibly interesting regions in the AC discrepancy maps in
Figures~\ref{fig:acdmap_comm15}--\ref{fig:acdmap_smica18} do appear to be basically local, especially as the biggest anomalies are in the
inpainted mask regions, where one would expect maximum distortion.

In this paper we assume that all the Fourier coefficients $a_{\ell,m}$
for $m = -\ell,..., \ell$, $\ell \leq L$ are available or can be readily calculated,
as is the case for the \textit{Planck} full-sky maps.
Physically, the map information is only reliable outside masked regions (see Section~\ref{sec:non-inpainted}),
but in this case only ``pseudo'' $C_\ell$ (see \cite{PhysRevD.64.083003} for example)
are available. A generalisation of the probe to be applicable to partial or masked sky maps will require further work.


We are aware of other work on directional statistics applied to CMB maps,
collected from the WMAP seven-year Internal Linear Combination
(ILC7) data by \cite{Naselsky_2012}, and from {\it Planck} 2015 data by
\cite{PhysRevD.89.023010, CHENG2016445}, but their purposes and methods are quite different from the current study.

The rest of the paper is organised as follows. In Section~\ref{sec:math.background} we review the necessary mathematical background for random fields on the unit sphere and define autocorrelations and AC discrepancy of scaled probe
coefficients.  In Section~\ref{sec:results} we report empirical results
for autocorrelations of the CMB maps. In Section~\ref{sec:randfield} we
generate realisations of isotropic Gaussian  random fields with best-fit
angular power spectrum from the {\it Planck} data, in order to demonstrate that these realisations exhibit nothing like the localised artefacts seen in the {\it Planck} data. In Section~\ref{sec:compareL}, we examine the effect of increasing cutoff multipole $L$ from its initial value of 1500 to a value of 2500 and observe that some interesting artefacts now appear. In Section~\ref{sec:non-inpainted} we consider the non-inpainted 2018 maps and then present a model to explain the most significant feature of the AC discrepancy map in the case of non-inpainted \texttt{SEVEM} 2018. Finally, in Section \ref{sec:conclusion} we give brief conclusions.

\section{Mathematical background}\label{sec:math.background}
In this section we recall some necessary mathematical background for
random and non-random fields on the unit sphere $\sph{2}$, and define
autocorrelation and autocorrelation discrepancy.

A real-valued field $T$ on the unit sphere can be expanded in terms of its Fourier series,
\begin{equation}\label{eq:Texp}
\begin{array}{l}
        \displaystyle T(\bx) = \sum_{\ell=0}^\infty\sum_{m=-\ell}^\ell a_{\ell,m}Y_{\ell,m}(\theta,\phi),\\[0.5cm]
\displaystyle a_{\ell,m}= \int_{\mathbb{S}^2} \overline{Y_{\ell,m}(\theta, \phi)} T(\bx) \intd{\bx},
\end{array}
\end{equation}
where the $Y_{\ell,m}(\theta,\phi)\equiv Y_{\ell,m}( \bx)$ for
$\ell=0,1,2,\ldots;\; m = -\ell,\ldots,\ell$ are the orthonormal
complex-valued spherical harmonics \citep{VarMosKhe88} of degree $\ell$,
and $\theta \in [0,\pi]$ is the polar angle of $\bx$ and $\phi \in
[0,2\pi)$ is the azimuthal angle. More explicitly,
\begin{align*}
 Y_{\ell,m} (\theta,\phi) &= \sqrt{ \frac{2\ell+1}{4\pi} \frac{(\ell-m)!}{(\ell+m)!} }
            P_{\ell,m} (\cos\theta) \exp( i m \phi), \quad m \ge 0, \\
 Y_{\ell,m} (\theta,\phi) &= (-1)^m \overline{Y_{\ell, -m}} (\theta, \phi), \quad m <0,
\end{align*}
where $P_{\ell,m}$ denotes the associated Legendre function, defined
in terms of the Legendre polynomial $P_\ell$ by
\[
 P_{\ell,m} (\mu) = (-1)^m (1-\mu^2)^{m/2} \frac{\mathrm{d}^m}{\mathrm{d}\mu^m} P_\ell(\mu),
\;\; m=0,1,\ldots,\ell
\]
for $\ell=0,1,2,\ldots$ and $\mu\in [-1,1]$.

As stated before, the probe coefficient for the field $T$ at degree
$\ell$ and direction $\bp$ is given by (see \eqref{eq:Tptobe} and
\eqref{eq:probedef})
\begin{equation}\label{eq:probecoeff}
T_{\ell,\bp}=
\sqrt{\frac{2\ell+1}{4\pi}}\int_{\mathbb{S}^2}
T(\bx)P_\ell(\bx\cdot\bp) \mathrm{d} \sigma(\bx).
\end{equation}
The probe coefficient is real, and has the symmetry shown in
\eqref{eq:symmTlp}, following immediately from
$P_{\ell}(-t)=(-1)^{\ell}P_{\ell}(t)$, $t\in [-1,1]$. The probe
coefficients are easily computed once the coefficients $a_{\ell,m}$ are
known, since from \eqref{eq:probecoeff} and \eqref{eq:Texp} we have
\begin{equation}\label{eq:probe_compute}
T_{\ell,\bp} = \sqrt{\frac{4\pi}{2\ell+1}}\sum_{m=-\ell}^\ell
a_{\ell,m}Y_{\ell,m}(\bp).
\end{equation}

When $T$ is a random field, following \cite[Chapter~5]{MaPe2011},
the field $T$ is
said to be \emph{strongly isotropic} if, for every $N \in \N$, every
$\bx_1, \ldots, \bx_N \in \Sp^2$ and every $\rho \in SO(3)$ (the group of
rotations in $\R^3$) the multivariate random vectors $\bigl(T(\bx_1), \ldots, T(\bx_N)\bigr)$ and $\bigl(T(\rho\bx_1), \ldots, T(\rho\bx_N)\bigr)$ have the same law.

Furthermore, $T$ is said to be \emph{Gaussian} if for all $N \in \N$ and for all $\bx_1,\ldots,\bx_N \in \Sp^2$ the random vector
$\bigl(T(\bx_1),\ldots,T(\bx_N)\bigr)$ has a  Gaussian distribution, i.e. $\sum_{j=1}^N a_j T(\bx_j)$ is a normally distributed random variable for all $a_j \in \R$, $j=1,\ldots,N$.

The expansion in \eqref{eq:Texp} is understood to converge in the sense that
\[
 \lim_{L\rightarrow \infty} \bE
\left[  \int_{\Sp^2} \left( T(\bx) - \sum_{\ell=0}^L \sum_{m=-\ell}^{\ell}a_{\ell,m} Y_{\ell,m} (\bx) \right)^2 \intd{\bx}
\right] = 0,
\]
and also for every fixed $\bx \in \Sp^2$,
\[
\lim_{L\rightarrow \infty}
\bE
\left[
  \left( T(\bx) - \sum_{\ell=0}^L \sum_{m=-\ell}^\ell a_{\ell,m} Y_{\ell,m}(\bx) \right)^2
\right] = 0.
\]

For an isotropic zero-mean random field $T$, the covariance function,
defined by  $G_{T}(\bx\cdot\bz):=\expect{T(\bx) \overline{T(\bz)}}$, is a
zonal function on $\mathbb{S}^2$, i.e. it depends only on the dot
product $\bx\cdot\bz$. It is given in terms of the angular power
spectrum $(C_\ell)_{\ell\geq2}$ by \eqref{eq:cov}.

If $T$ is an isotropic Gaussian random field then the probe coefficients
$T_{\ell,\bp}$ for any fixed $\bp$ and variable $\ell\ge 2$ are
uncorrelated mean-zero Gaussian random variables with variance $C_\ell$,
since from \eqref{eq:probe_compute} and \eqref{mean var alm} we have
\begin{align*}
\expect{T_{\ell,\bp}}&=0,\\
\expect{ T_{\ell,\bp} \overline{T_{\ell',\bp}}}
  &= \frac{4\pi}{2\ell+1}
\sum_{m=-\ell}^\ell
\sum_{m'=-\ell'}^{\ell'}
\Exp{a_{\ell,m} \overline{a_{\ell',m'}}}Y_{\ell,m} (\bp) \overline{Y_{\ell',m'}(\bp)}  \\
&=
   \frac{4\pi}{2\ell+1}
\sum_{m=-\ell}^\ell
\sum_{m'=-\ell'}^{\ell'}
           C_\ell \delta_{\ell,\ell'} \delta_{m,m'} Y_{\ell,m} (\bp)
   \overline{Y_{\ell',m'}(\bp)} \\
&= \begin{cases}
   0,  & \mbox{ if } \ell \ne \ell', \\
 \displaystyle \frac{4\pi}{2\ell+1} C_\ell  \sum_{m=-\ell}^\ell Y_{\ell,m} (\bp) \overline{Y_{\ell,m}(\bp)}
= C_\ell, & \mbox{ if } \ell = \ell',
   \end{cases}
\end{align*}
where in the last step we used the addition theorem~\eqref{addition} with $\bx=\bz=\bp$.

For a given map and fixed $\bp$, we explore the correlations between
the successive values of the scaled $T_{\ell,\bp}$  for varying $\ell$ by
making use of the notion of \emph{autocorrelation}. Autocorrelation plots
\citep{BoJeReLj2015} are a commonly-used tool for studying correlation in a time series, in which case the autocorrelations are computed for
different time lags. In the present case, the autocorrelation is between
scaled values of $T_{\ell,\bp}$ for a fixed direction $\bp$ and variable
$\ell$ and the lags are with respect to values of $\ell$, not time.
(Scaling is necessary so that each supposedly independent random variable has the same variance.) If the scaled coefficients are iid standard normal random variables, such autocorrelations mostly lie
within the $95.45\%$ confidence intervals (i.e., the blue boundary lines)
for any $\bp$ and all positive lags, as in Figure
\ref{fig:comm15_autocorr}. Autocorrelations which display distinct
patterns, as in Figure~\ref{fig:comm15_ac_maxacd}, are certainly not compatible with realisations of iid random variables.

Let $\PT{q}:=\{q_{\ell}\}_{\ell=2}^{L}$, $L\in\N$, be a finite
real-valued sequence. (We start from $\ell=2$ because the
CMB maps have the monopole and dipole subtracted.) Let
\begin{equation}\label{eq:mean}
        \widehat{\PT{q}}_L := \frac{1}{L-1}\sum_{\ell=2}^{L}q_{\ell}
\end{equation}
be the empirical mean of $\PT{q}$. For $k=0,1,\dots,k_{\max}$, let
\begin{equation}\label{eq:betadef}
   \beta_{k} := \frac{1}{L-1}\sum_{\ell=2}^{L-k}\bigl(q_{\ell}-
   \widehat{\PT{q}}_L\bigr)\bigl(q_{\ell+k}-\widehat{\PT{q}}_L\bigr).
\end{equation}
The \emph{autocorrelation} of $\PT{q}$ for lag $k$ is
\begin{equation}\label{eq:ACdef}
        \alpha_{k} := \alpha_{k}(\PT{q}) :=\frac{\beta_{k}}{\beta_0}.
\end{equation}
See e.g. \cite{BoJeReLj2015} and \cite{Hamilton1994}.

To apply this to the probe coefficients $T_{\ell,\bp}$, we first
scale them by dividing by the square root of the (empirical) variance $\widehat{C}_\ell$ given by \eqref{eq:empiricalC}.
Given a probe direction $\bp\in\sph{2}$ and $L\in\N$ the scaled probe coefficients $\widetilde{T}_{\ell,\bp}$ defined in \eqref{eq:wtdT} should, to a good approximation, be instances of independent standard normal random variables, and their autocorrelations should be uncorrelated.

To understand the close relation between the autocorrelations at
$\bp$ and $-\bp$, as foreshadowed in the Introduction, we note that if (as is very often the case) the mean defined by \eqref{eq:mean} is small
enough to be neglected, then from \eqref{eq:betadef}
using \eqref{eq:symmTlp} we see that $\beta_k(-\bp) = (-1)^k \beta_k(\bp)$, from which it follows using \eqref{eq:ACdef} that the autocorrelations at $\bp$ and $-\bp$ differ by the same factor $(-1)^k$.

We have already seen in Figures~\ref{fig:comm15_autocorr} and \ref{fig:comm15_ac_maxacd} the graphs of autocorrelations of scaled probe coefficients $\widetilde{T}_{\ell,\bp}$ for \texttt{Commander} 2015, for two different directions $\bp$. In order to find other directions
$\bp$ in which the scaled probe coefficients have significant departures
from randomness, we introduce a measure of that departure, the
\emph{autocorrelation discrepancy}, or \emph{AC discrepancy}:
\begin{definition}\label{def:ACD}
For a field $T$ on the sphere, the \emph{autocorrelation discrepancy with maximum lag $k_\mathrm{max}$} is
\begin{equation*}
        \discrep_{k_\mathrm{max}}(\bp):=\sum_{k=1}^{k_\mathrm{max}}
\max\left\{|\alpha_{k}(\widetilde{T}_{\cdot,\bp})|-t_{L},0\right\},
\end{equation*}
where the autocorrelation $\alpha_{k}$ is given by \eqref{eq:ACdef}, the
sequence $\widetilde{T}_{\cdot,\bp}$ is given by \eqref{eq:wtdT}, and
$t_{L}$ is the \emph{threshold constant} given by
\eqref{eq:threshold}.
\end{definition}

The function $D_{k_\mathrm{max}}(\bp)$ with maximum lag $k_\mathrm{max}$ and cutoff multipole $L$ and varying direction $\bp$ is called for short the \emph{AC discrepancy map} for the field $T$.

For iid data with finite variance (as would be expected for a Gaussian CMB map), the autocorrelations $\alpha_{k}$, $k>0$, are approximately iid Gaussian random variables with mean $0$ and a lag-dependent variance of approximately $(L-k-1)/(L-1)^2$~\citep{PfeiferDeutsch:1981}. Given a lag-independent threshold constant $t_L = 2/\sqrt{L-1}$, the probability of an autocorrelation of lag $k$ falling within the range
$[-t_L,t_L]$ is thus given by $\mathbb{P}(|\alpha_k| \leq t_L) \simeq
\mathrm{erf} \left[ \sqrt{2} \sqrt{L-1}/\sqrt{L-k-1}\right]$. Hence, for
a Gaussian map, the probability of the AC discrepancy with maximum
lag~$k_\mathrm{max}$ being zero for the given direction (which
is equal to the probability that $|\alpha_k| \leq t_L$ for all $k = 1,\ldots, k_{\mathrm{max}}$) is
\begin{equation}
\label{eq:zeroprob}
\mathbb{P}(D_{k_\mathrm{max}} = 0) \simeq  \prod_{k = 1}^{k_\mathrm{max}} \; \mathrm{erf} \left[ \sqrt{2} \sqrt{L-1}/\sqrt{L-k-1}\right].
\end{equation}

\section{Empirical CMB autocorrelations}\label{sec:results}
In this section, we will apply the probe to the
 CMB maps produced by the {\it Planck} consortium, to obtain the AC discrepancy results shown in Figures \ref{fig:acdmap_comm15}, \ref{fig:acdmap_comm18}, \ref{fig:acdmap_nilc18}, \ref{fig:acdmap_sevem18} and \ref{fig:acdmap_smica18}.

 The temperature maps themselves are obtained by different component separation methods:
\begin{itemize}
\item \texttt{Commander} \citep{Er_etal2006,Er_etal2008} is a Bayesian
    parametric method that works in the map domain.
\item \texttt{NILC} \citep{De_etal2009} is an implementation of an internal
    linear combination (ILC) that works in the needlet domain.
\item \texttt{SEVEM} \citep{Fe_etal2012} is an implementation of the
    template-cleaning approach to component separation that works in
    the map domain. Foreground templates are constructed by
    differencing pairs of maps from the low- and high-frequency
    channels.
\item \texttt{SMICA} \citep{Ca_etal2008} is a non-parametric method that works in the spherical harmonic domain.
\end{itemize}

In the experiments we take $L=1500$ and use the data at the
$50,331,648$ \texttt{HEALPix} points ($N_\textrm{Side} = 2048$).
The \texttt{healpy} package\footnote{\url{http://healpix.sourceforge.net}}
\citep{Gorski_etal2005} is used to calculate the Fourier coefficients
$a_{\ell,m}$. In Figures~\ref{fig:acdmap_comm15}, \ref{fig:acdmap_comm18}, \ref{fig:acdmap_nilc18}, \ref{fig:acdmap_sevem18} and \ref{fig:acdmap_smica18} we probe over {$12,582,912$} \texttt{HEALPix} points $\bp$ (corresponding to $N_{\textrm{Side}}=1024$) to produce the AC discrepancy maps $D_{k}(\bp)$ with maximum lag $k_{\max}=10$. A reasonable observation is that some of the five AC discrepancy maps (especially Commander 2018, SEVEM and SMICA) have significant departures from the assumed model of an isotropic Gaussian random field, especially in the masked regions of the CMB maps.

\section{Random fields}\label{sec:randfield}
In this section, we study the properties of isotropic Gaussian random
fields on the sphere, so we can exclude the possibility that the AC
discrepancies seen in some of Figures \ref{fig:acdmap_comm15} to
\ref{fig:acdmap_smica18} are accidental realisations of such a random
process.

\begin{figure}
\centering
\includegraphics[trim = 0mm 0mm 0mm 0mm, width=0.75\columnwidth]{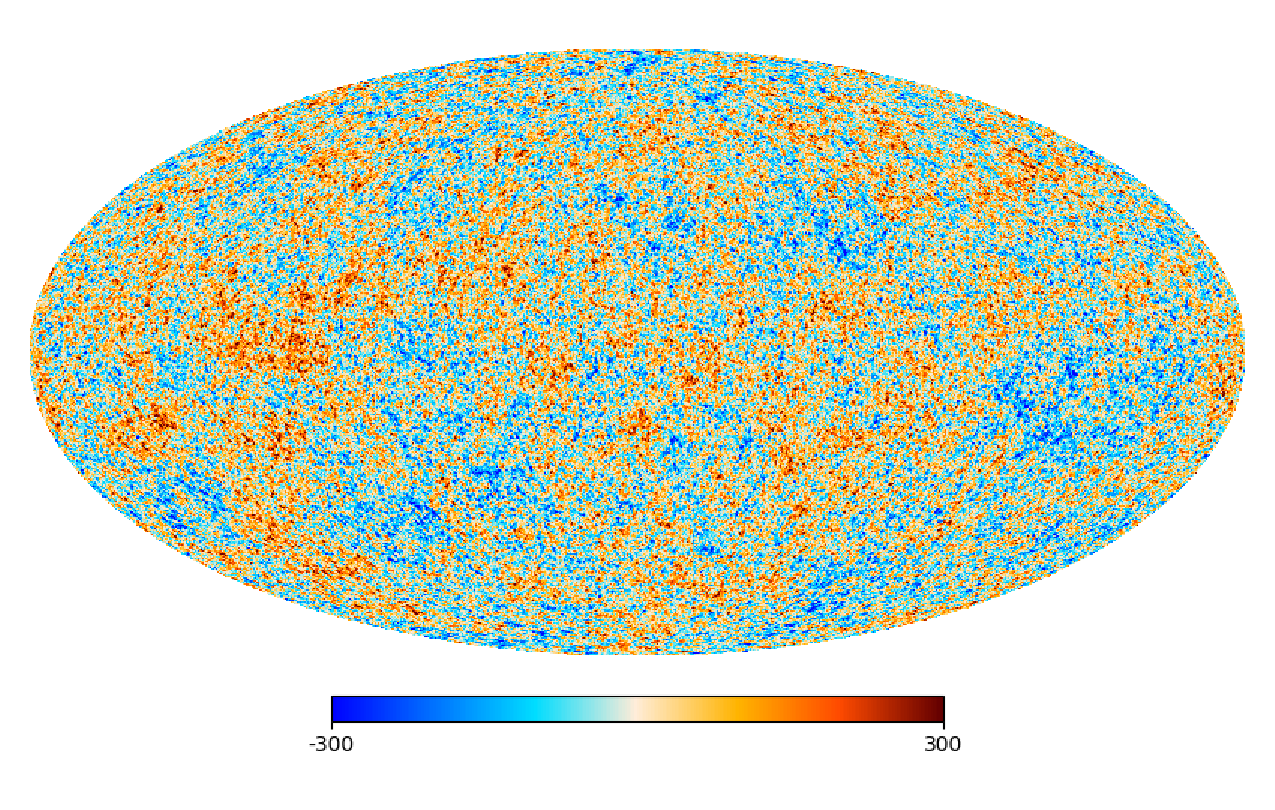}\vspace{-5mm}
\caption{Instance $1$ of Gaussian random field generated from the best-fit angular power spectrum up to degree $2508$, $N_{\textrm{Side}} = 2048$}
\label{fig:grf_instance1}
\vspace{3mm}
\includegraphics[trim = 0mm 0mm 0mm 0mm, width=0.75\columnwidth]{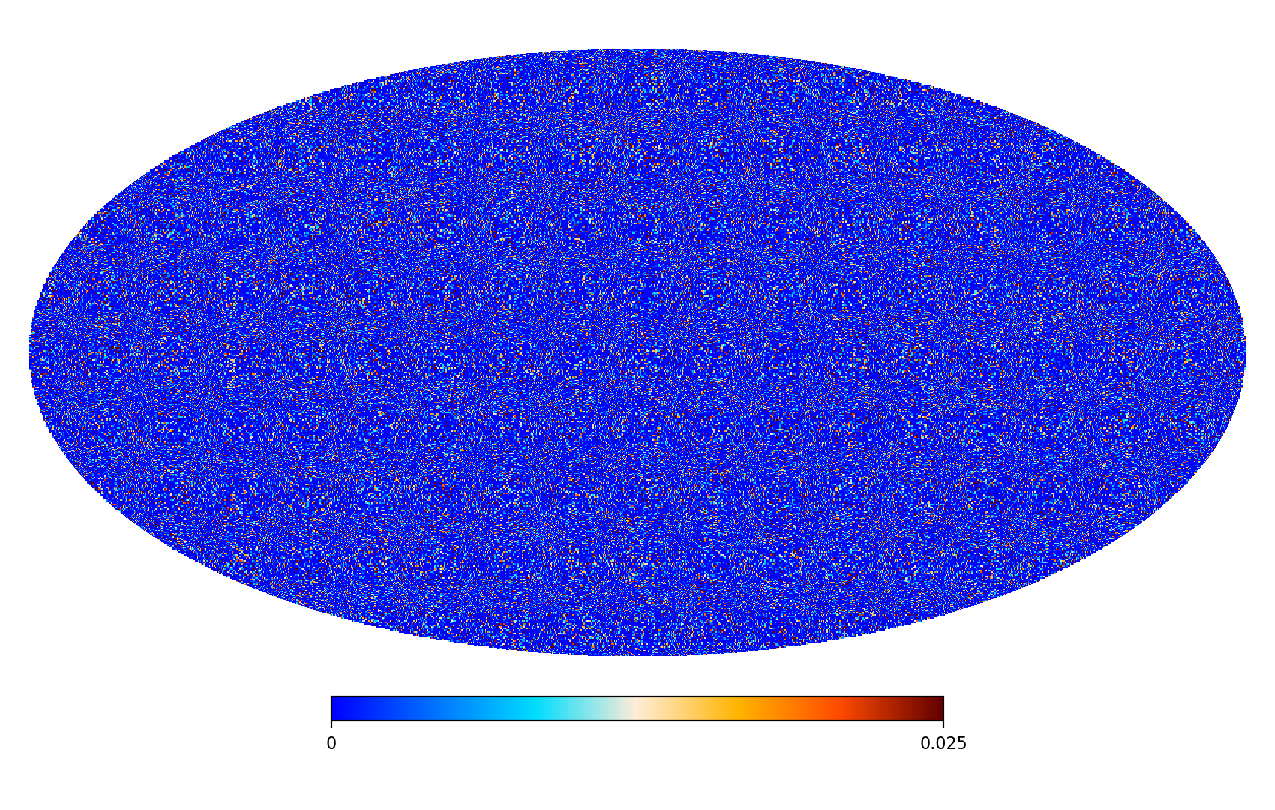}\\[-4mm]
\caption{AC discrepancy map, instance $1$ of Gaussian Random Field in Figure~\ref{fig:grf_instance1}, $N_{\textrm{Side}} = 1024$, $k_{\max} = 10$, ${\rm max} = 0.147$ at $(l,b) = (132.70, -64.71)$, $L=1500$}
\label{fig:acd_instance1}
\end{figure}

\begin{figure}
\begin{minipage}{\columnwidth}
        \centering
\includegraphics[width=0.75\columnwidth]{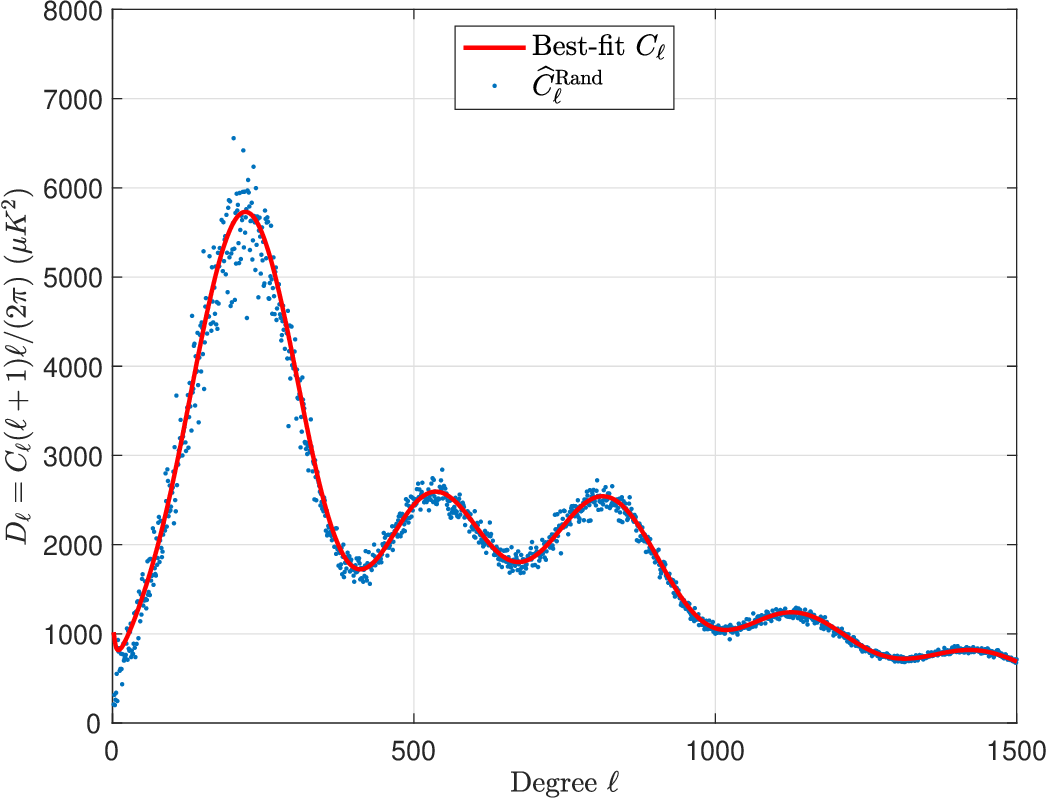}\\
\caption{Angular power spectrum, instance $1$ of Gaussian Random Field in Figure~\ref{fig:grf_instance1}}
\label{fig:aps_instance1}
\end{minipage}
\end{figure}

First, in Figure~\ref{fig:grf_instance1} we show one realisation of an isotropic Gaussian random field with the best-fit angular power spectrum provided by \cite{Planck2018I} at $N_{\rm Side}=2048$ using the \texttt{healpy} package. For this realisation, in Figure~\ref{fig:acd_instance1} we show its AC discrepancy map, and
in Figure~\ref{fig:aps_instance1} we show the empirical angular power spectrum.
We observe in the AC discrepancy map a pattern of apparently uniformly distributed random small discrepancies
over the whole sphere, as is expected.
%
To strengthen that conclusion we computed in total $10$ independent
realisations similar to Figure~\ref{fig:grf_instance1}, and in every case
found a similarly uninteresting AC discrepancy map.

A useful statistic to test the assumption of isotropic Gaussian random fields is the proportion of zero pixels in the AC discrepancy maps, which we will call $f_0$. As explained already, the probe coefficients $(T_{\ell,\bp})_{\ell=2}^L$ for a given direction $\bp$ should be realisations of independent Gaussian random variables with variance $C_\ell$. The scaled probe coefficients $\Bigl(T_{\ell,\bp}/\sqrt{\widehat{C}_\ell}\Bigr)_{\ell=2}^L$ should, therefore, be independent samples from a standard normal distribution,
at least to the extent that the empirical variances are close to the true variances. It is well known (see for example {\cite[p.16]{BroDav2016}})
that their autocorrelations $(\alpha_{k})_{k=1 \ldots 10}$  should also be approximately independent for large $L$.  More precisely, the autocorrelations for different lags $k$ are themselves uncorrelated, since it can be seen that in (\ref{eq:betadef}) every product of the mean-zero random variables $q_\ell$, $q_{\ell + k}$, $q_{\ell'}$, $q_{\ell'+ k'}$, at least one of the subscripts is different from all the others.  From the uncorrelation and the Gaussianity of the underlying variables, the autocorrelations are asymptotically independent \cite{PecTud2005}.  Given the large number $L-1 = 1499$ of variables, in this case, it would be reasonable to assume independence of the autocorrelations to sufficient accuracy.

 As per Equation~\eqref{eq:zeroprob}, the probability of the AC discrepancy with maximum lag \mbox{$k_\mathrm{max} = 10$} being zero is
$\mathbb{P}(D_{10} = 0) \approx 0.6303$. For comparison, using our sample of $10$ random field realisations, we find the fraction of the pixels in the AC discrepancy map having the value zero is ${\widehat{f}}_0^{\rm ran} = 0.6315 \pm 0.0004$, which is reasonably close to the analytically predicted value of \mbox{$f_0^{\rm ran} = \mathbb{P}(D_{10} = 0) \approx 0.6303$}, bearing in mind that the independence is only approximate, and the fact that we are using empirical rather than true variances for the probe coefficients. The AC discrepancy maps of the {\it Planck} maps, on the other hand, have slightly lower fractions of zeros, ranging from $f_0 = 0.5996$ (\texttt{SEVEM}~2018) to $0.6010$
(\texttt{Commander}~2018), $0.6204$ (\texttt{SMICA}~2018) and $0.6236$ (\texttt{NILC}~2018) to $0.6289$ (\texttt{Commander}~2015).
Note that for a single full-sky realisation of a map, we statistically expect a deviation from the ensemble average, which can be estimated in terms of the standard deviation of $f_0$ in the simulations.  For our 10 simulations, this standard deviation is ${\widehat{\sigma}}_{f_0} \sim 0.0013$. Thus of the inpainted maps only \texttt{Commander}~2015 has $f_0$ consistent with the hypothesis of an isotropic Gaussian random field.

We also investigate the expected impact of CMB lensing on the AC discrepancy maps by generating lensed versions of
our random maps using the \texttt{lenspyx} CMB lensing package\footnote{\url{https://github.com/carronj/lenspyx}},
and evaluating the corresponding AC discrepancy maps.  Averaging over the differences between $f_0$ for lensed and unlensed versions of the same random realisation, we find that lensing decreases the expected fraction of zeros in the AC discrepancy map by ${\widehat{\Delta f_0}} = -0.00042 \pm 0.00026$.  It is smaller than {$\widehat{\sigma}_{f_0}$}, so for a single map, the AC discrepancy will not be able to distinguish the lensing effect from sample variance.  Moreover, we verified that pixelisation effects related to the transformation of maps from harmonic to pixel space and back do not affect our estimates of $f_0$. That is, the difference between $f_0$ for a random map given in harmonic space and the same map transformed to pixel space and back via \texttt{HEALPix}'s \texttt{alm2map} and $\texttt{map2alm}$ routines is ${\widehat{\Delta f_0}} = \mathcal{O}(10^{-7})$, i.e., completely negligible.

For a more quantitative analysis of CMB maps, it would be straightforward to use simulations of Gaussian random fields like the ones considered in this section to construct a likelihood function, based on, e.g., $f_0$, or another observable, such as a suitable binning of the AC discrepancy's one-point function. 

\section{Comparing the ACD maps for $L=1500$ and $L=2500$}\label{sec:compareL}
We recall that in Section~\ref{sec:math.background}, the probe
coefficients $T_{\ell,{\bf p}}$ are enumerated from $\ell=2$ to $\ell=L$,
where $\ell$ is the degree of the Legendre polynomial used in the probe
\eqref{eq:probecoeff}. The higher the degree $\ell$, the more oscillatory
the polynomial $P_{\ell}$ becomes. So intuitively, the value $L$ indicates
the finest resolution of the field $T$ that the probe can analyse. With a
given set of scaled probe coefficients $(T_{\ell,{\bf
p}}/\sqrt{\widehat{C}_{\ell}})$, for $\ell=2,\ldots,L$, the
autocorrelation discrepancy map (or ACD map) was computed using the
formula given in Definition~\ref{def:ACD}.

\begin{figure}[h]
\centering
\includegraphics[trim = 0mm 0mm 0mm 0mm, width=0.75\columnwidth]{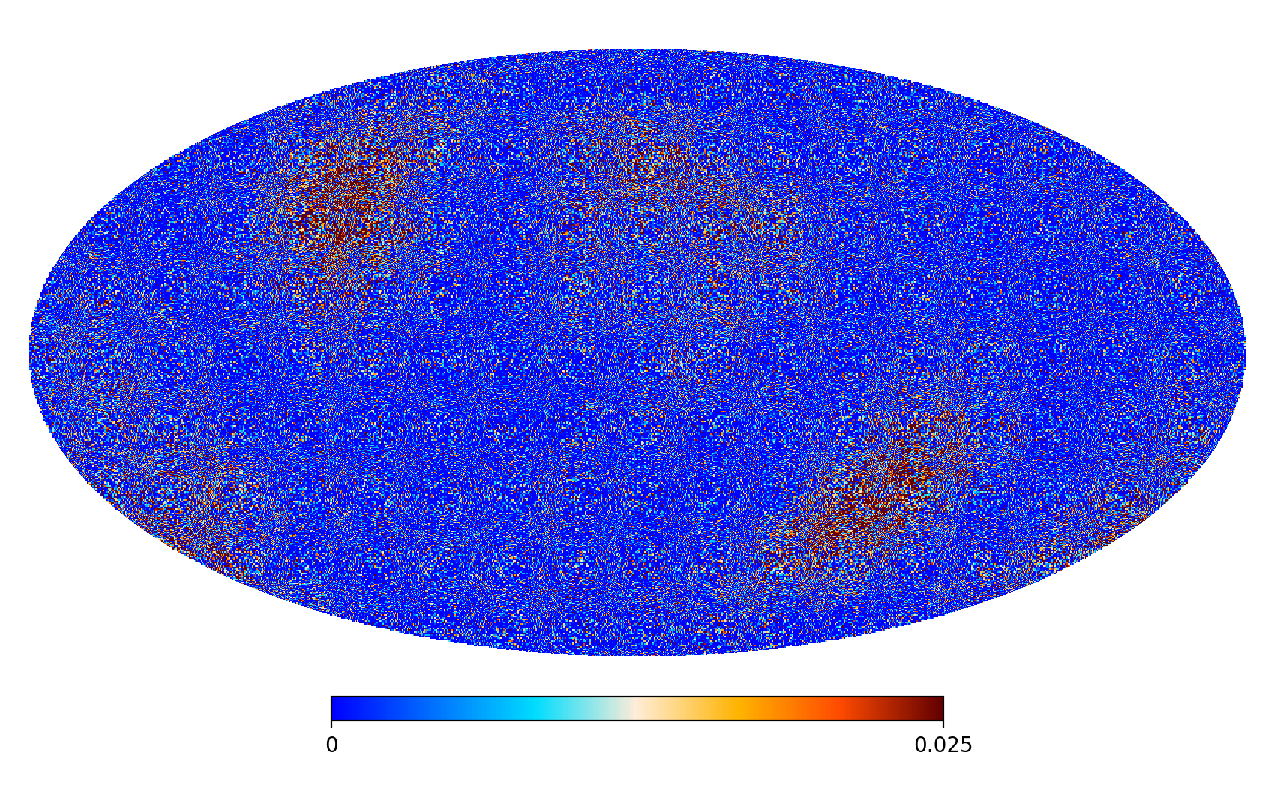}\\[-4mm]
\caption{AC discrepancy map for L = 2500, \texttt{Commander} 2015, $N_{\textrm{Side}} = 1024$,
$k_{\max} = 10$, ${\rm max} = 2.075$ at $(l,b) = (35.20, 24.30)$}
\label{fig:acdmap_comm15_L2500}\vspace{3mm}
\end{figure}

\begin{figure}[ht]
\vspace{-6mm}
\centering
\includegraphics[trim = 0mm 0cm 0mm 0cm, width=0.75\columnwidth]{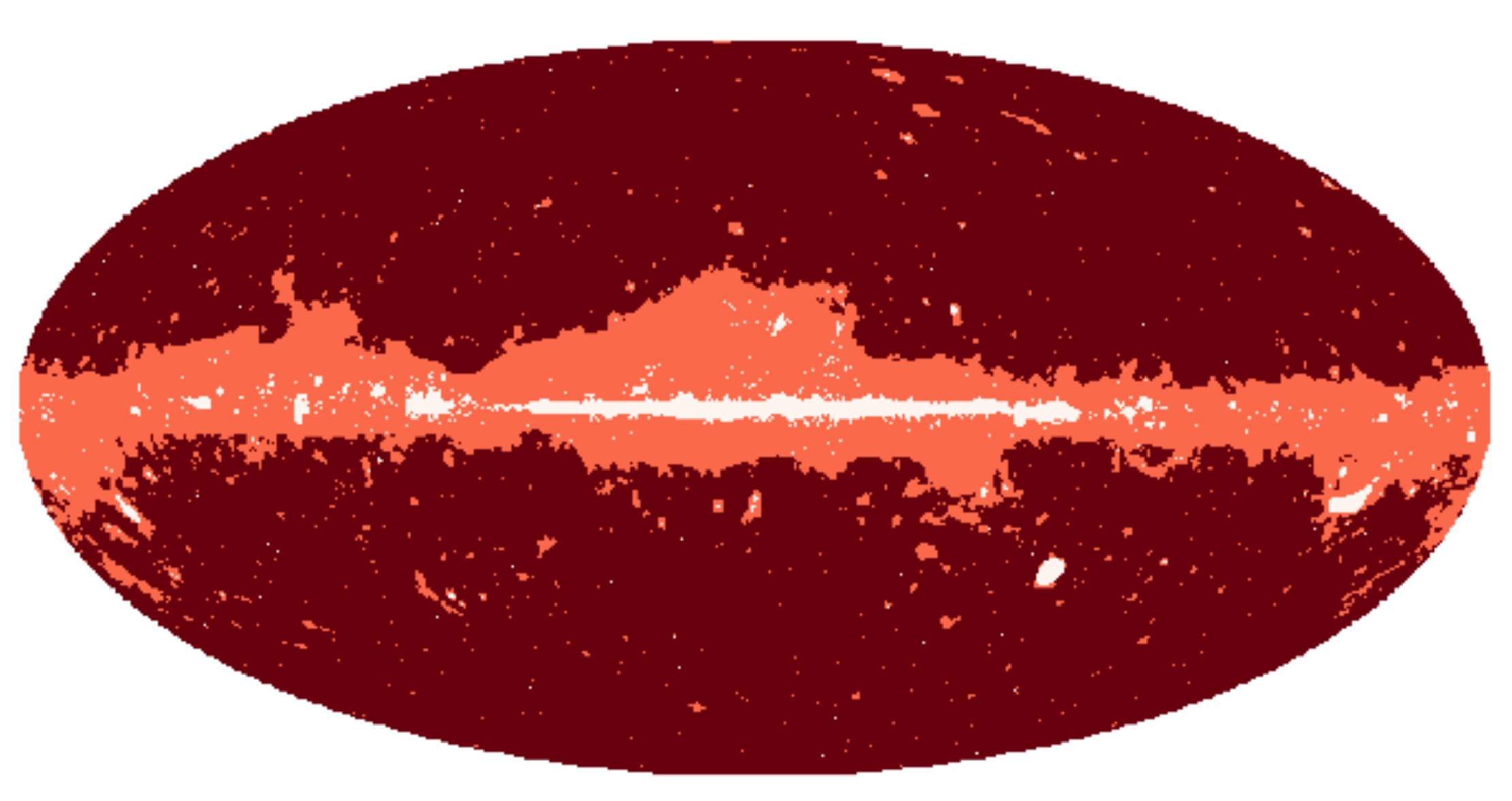}\\[-1mm]
\caption{\label{fig:mask2018}
\textit{Planck} temperature inpainting (white) and common confidence masks (light red)~\cite{planck-wiki-masks}.}
\end{figure}

In this section, we investigate what happens when the value of
$L$ is increased. The most interesting change is that in the ACD map
for Commander 2015 with $L=2500$ in Fig~\ref{fig:acdmap_comm15_L2500}, we see two ``blobs'' (at two antipodes) while with the same colour map these features are not present in Fig~\ref{fig:acdmap_comm15}, which is the ACD map for the same Commander 2015 temperature map with $L=1500$.
The phenomenon does not occur for other CMB maps when their ACD
maps with $L=2500$ are computed, except that the size of the departure from uniformity near the galactic plane all become considerably larger. Curiously, the ``blobs'' in Fig~\ref{fig:acdmap_comm15_L2500} appear far from the galactic plane, in regions that lie outside the \textit{Planck} temperature inpainting and confidence masks (cf. Figure~\ref{fig:mask2018}). However, since they are
neither reproduced by the other component separation methods nor survive in the 2018 iteration of \texttt{Commander}, it seems highly unlikely that they point to an underlying physical effect.
On the other hand, they do point to some anomalous short-range
information present in the Commander 2015 temperature map.

\section{The non-inpainted 2018 maps}\label{sec:non-inpainted}
\begin{figure}
\centering
\includegraphics[trim = 0mm 0mm 0mm 0mm, width=0.75\columnwidth]{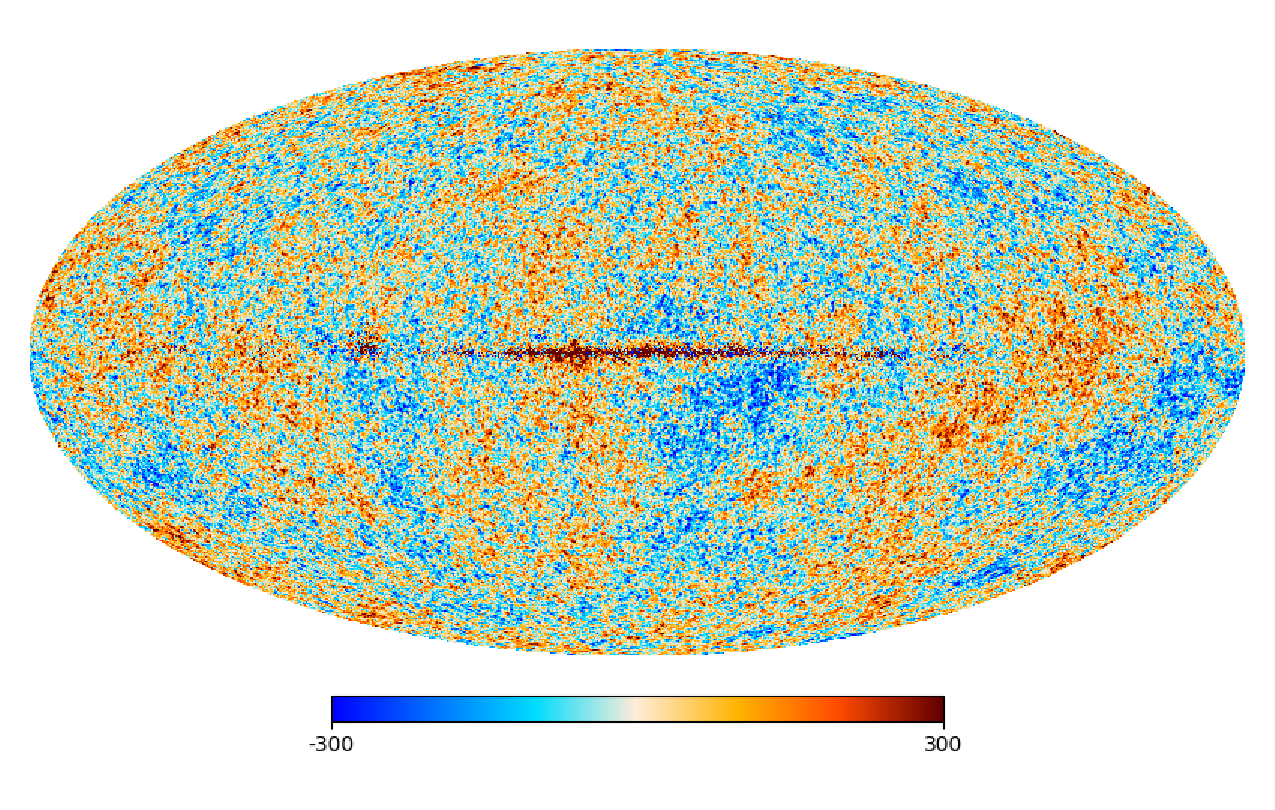}\\[-4mm]
\caption{Non-inpainted CMB map of \texttt{SEVEM} 2018}
\label{fig:cmb_UIsevem2018}
\vskip 2mm
\includegraphics[trim = 0mm 0mm 0mm 0mm, width=0.75\columnwidth]{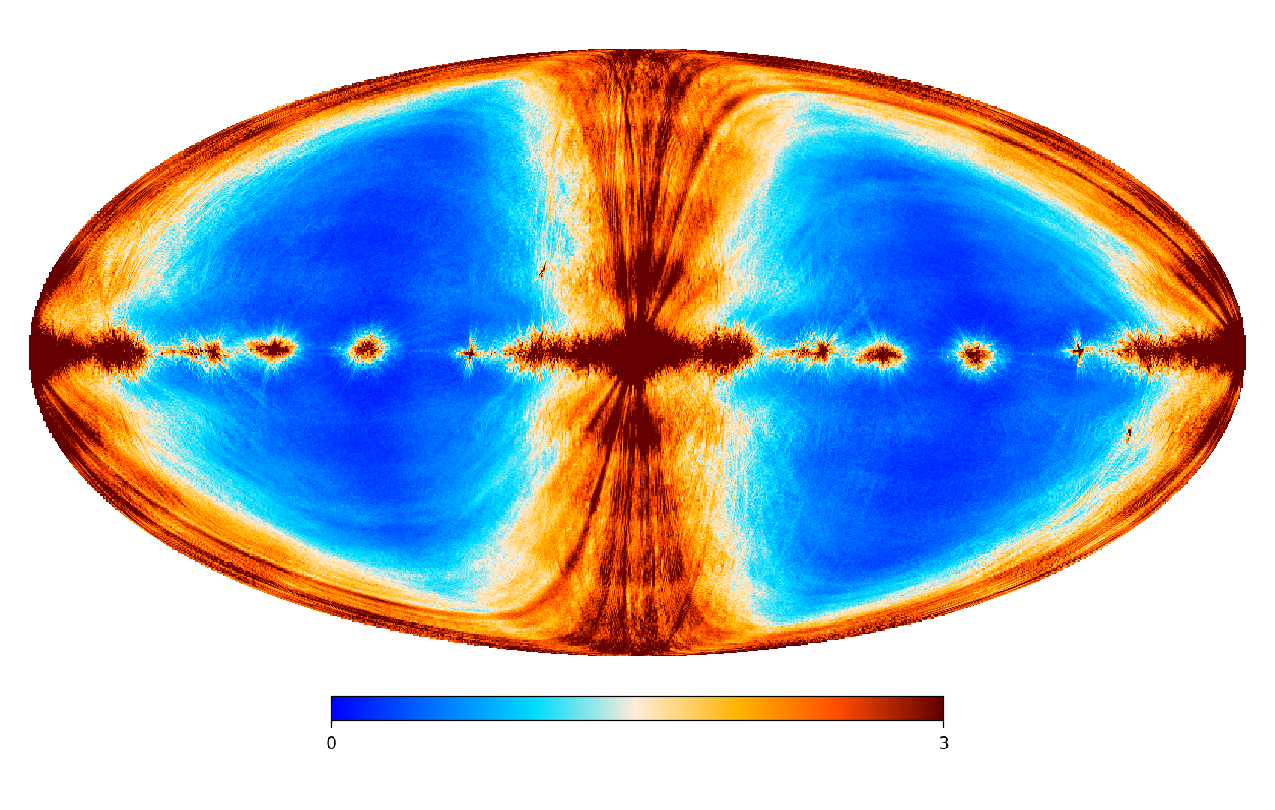}\\[-5mm]
\caption{AC discrepancy map for non-inpainted CMB map of \texttt{SEVEM} 2018, $L = 2500$,
$N_{\textrm{Side}} = 1024$,
$k_{\max} = 10$, ${\rm max} = 9.37$ at $(l,b) = (180.62, 0.04)$}
\label{fig:acd_UIsevem2018}
\end{figure}

To this point, we have considered only the inpainted {\it Planck}
temperature maps. In this section, in contrast, we consider the
non-inpainted {\it Planck} 2018 temperature maps. At face value they have less interest to us because they generally exhibit obvious pollution near the galactic equatorial plane; for the case of non-inpainted
\texttt{SEVEM} 2018 see Figure~\ref{fig:cmb_UIsevem2018}. Such
equatorial pollution is clearly not consistent with the field being a realisation
of a Gaussian random field. Nevertheless, there is value in applying our
probe in such cases because (as we see in Figure~\ref{fig:acd_UIsevem2018} in the case of \texttt{SEVEM} 2018) the AC discrepancy maps have a very different character from that seen in Figures~\ref{fig:acdmap_comm15}--\ref{fig:acdmap_smica18} for the inpainted maps. The magnitudes in the AC discrepancy map
Figure~\ref{fig:acd_UIsevem2018} are firstly much larger: note that the
colour map is many times larger in scale than those in earlier maps,
and even then is saturated. Secondly, the high values occur globally,
especially on the great circle through the galactic centre and the poles.
Thirdly, in contrast to the apparent white noise character of the AC
discrepancies for the inpainted maps, the AC discrepancies now seem to be continuous.

We now present a model field with the same principal characteristics for the AC discrepancy as Figure~\ref{fig:acd_UIsevem2018}, yet which consists simply of an
isotropic Gaussian random field plus a narrow needlet-like structure at
the galactic centre. The lesson to be learned is that the AC discrepancy can be a non-local rather than a local function of artefacts in the temperature maps.

\begin{figure}
\centering
\begin{minipage}{0.75\textwidth}
\centering
    \includegraphics[trim = -1.6cm 0mm 0mm 0mm, width=\textwidth]{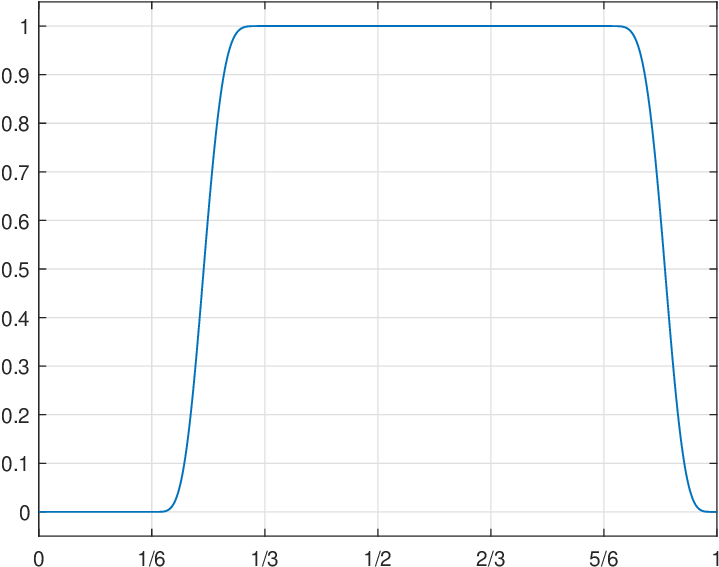}\\[5mm]
\caption{Plot of the function $h$ defined in~\eqref{def:h}}
\label{fig:function h}
\end{minipage}
\end{figure}

Our model in this case takes the form
\begin{equation}\label{eq:T total}
T^{\rm total}(\bx) = T^{\rm{rand}}(\bx) + \gamma T^{\rm{need}}(\bx),
\end{equation}
where $\gamma$ is an adjustable parameter,
$T^{\rm{rand}}$ is a mean-zero, strongly isotropic Gaussian random field with angular power
spectrum $(C_\ell)_{\ell\ge 2}$ as in Section~\ref{sec:randfield},
and $T^{\rm{need}}$ is a narrow ``needlet-like'' spherical polynomial
of (high) degree $L_0$, of the general form: for $\bx_0,\bx\in\sph{2}$,
\begin{equation}\label{eq:dineed}
{T^{\rm{need}}(\bx)}
:= \sum_{\ell=2}^{L_0}H_{\ell}\sum_{m=-\ell}^{\ell}G_{\ell,m}\shY\bigl(\rho^{-1}\bx\bigr)\overline{\shY\bigl(\rho^{-1}\bx_0\bigr)},
\end{equation}
where $\rho$ is an arbitrary rotation matrix in $SO(3)$, and $H_{\ell}$ and $G_{\ell,m}$
are ``filters'' satisfying
\begin{equation*}
        0\leq H_{\ell}\leq 1,\quad 0\leq G_{\ell,m}=G_{\ell,-m}\leq1\mbox{~for~} |m|\leq \ell\leq L_0.
\end{equation*}

We shall say that the function defined by \eqref{eq:dineed} is a needlet-like field centred at $\bx_0$. In our applications, $\bx_0$ is the galactic centre with $(l,b) = (0,0)$.

From \eqref{eq:dineed}, the Fourier coefficient
$a^{\rm{need}}_{\ell,m}$ of $T^{\rm{need}}$ (with general $G_{\ell,m}$) is given by
\[
  a^{\rm{need}}_{\ell,m} = H_{\ell} G_{\ell,m} \overline{Y_{\ell,m} (\rho^{-1}\bx_0)}.
\]

The simplest case is that in which $G_{\ell,m}=1$ for all $\ell,m$. In that case, the addition theorem \eqref{addition} reduces the right-hand side of \eqref{eq:dineed} to
\begin{equation*}
        \frac{1}{4\pi} \sum_{\ell=2}^{L_0}(2\ell+1)H_{\ell}P_{\ell}\bigl((\rho^{-1}\bx)\cdot(\rho^{-1}\bx_0)\bigr)
        = \frac{1}{4\pi} \sum_{\ell=2}^{L_0}(2\ell+1)H_{\ell}P_{\ell}\bigl(\bx\cdot\bx_0\bigr),
\end{equation*}
since the dot product is invariant under rotation. Thus in
the case $G_{\ell,m} \equiv 1$ our needlet-like field is rotationally symmetric about $\bx_0$.
A true needlet (see \cite{NarPetWar06,MarPie_etal,WaLeSlWo2017,LeSlWaWo2017}) has $G_{\ell,m}=1$ for all $\ell,m$ and
\begin{equation}\label{eq:Hell}
        H_{\ell}=h\left(\frac{\ell}{L_0}\right),
\end{equation}
where $h$ is a smooth function with support in $(0,1)$. The smoothness
and the compactness of the support ensures that the needlet decays rapidly for $\bx$ away from $\bx_0$.

In the present work we take $H_\ell$ to be of the form~\eqref{eq:Hell},
with $L_0 = \tfrac{6}{5}L$ and
\begin{equation}\label{def:h}
h(t) =
\begin{cases}
  p  (2-6t) &  \text{ for } t \in [1/6, 1/3], \\
  1       &  \text{ for } t \in [1/3, 5/6], \\
  p(6t-5) &  \text{ for } t \in [5/6, 1], \\
  0       & \text{ otherwise},
\end{cases}
\end{equation}
with $p(t) =924(1-t)^6  -4752(1-t)^7 + 10395(1-t)^8-12320(1-t)^9 +
8316(1-t)^{10}-3024(1-t)^{11}+462(1-t)^{12}$, see \cite{WaLeSlWo2017}. We note that $h$, as illustrated in Figure~\ref{fig:function h}, and its first
five derivatives, are continuous, since $p^{(\kappa)}(0) = 0$, $\kappa =
1,\ldots,6$ and $p^{(\kappa)}(1) = 0$, $\kappa = 1,\ldots,5$, and
$p(0)=1$, $p(1)=0$.

From formula \eqref{eq:probe_compute}, when $G_{\ell,m}=1$, the
corresponding probe coefficient $T^{\rm{need}}_{\ell,\bp}$ for any
direction $\bp \in \Sp^2$ is given by
\begin{equation}\label{eq:Tlp_need}
   T^{\rm{need}}_{\ell,\bp}
      = \sqrt{\frac{2\ell+1}{4\pi}}
            H_{\ell} P_{\ell}(\bx_0 \cdot \bp),
\end{equation}
where we again used the addition theorem \eqref{addition}.

To determine the autocorrelation of the probe coefficients of the
needlet, they must first be scaled. We do this for the case $G_{\ell,m}=1$ by dividing the non-zero probe coefficients by $\sqrt{\widehat{C}^{\rm{need}}_\ell}$, where
\begin{equation}\label{eq:Cl_need}
\widehat{C}^{\rm{need}}_\ell := \frac{1}{2\ell+1}\sum_{m=-\ell}^\ell
\left|a^{\rm{need}}_{\ell,m}\right|^2
=\frac{1}{2\ell+1}\sum_{m=-\ell}^\ell
H_{\ell}^2\left|Y_{\ell,m}(\bx_0)\right|^2
=\frac{1}{4\pi} H_{\ell}^2.
\end{equation}
Thus our input to the autocorrelation algorithm is in this case
\begin{equation}\label{eq:qLneed}
q^{\rm{need}}_\ell :=
\frac{T^{\rm{need}}_{\ell,\bp}}{\sqrt{\widehat{C}^{\rm{need}}_\ell}}
= \sqrt{2\ell+1}P_\ell(\bx_0\cdot\bp),\quad L_0/6 \le \ell \le L.
\end{equation}
We now use the (Laplace) asymptotic expression for the Legendre polynomial
\cite[p.194]{Szego1975}: for $\lambda\in(0,\pi)$,
\[
P_\ell(\cos \lambda)
= \sqrt{2}\left(\pi \ell \sin {\lambda}\right)^{-1/2}
\cos\left((\ell+\tfrac{1}{2}){\lambda} -\pi/4\right)+O\left(\ell^{-3/2}\right).
\]
Then from \eqref{eq:mean} we have, with $\cos\lambda=\bp\cdot\bx_0 \in (0,1)$
\begin{align*}
\widehat{q}^{\rm{need}}_L &=
  \frac{1}{L-1}  \sum_{\ell={L_0/6}}^L q^{\rm{need}}_\ell \\
  & = \frac{1}{L-1} \frac{2}{\sqrt{\pi}} \frac{1} {\sqrt{\sin \lambda}}
    \sum_{\ell=L_0/6}^L
   {\cos\bigl((\ell+\tfrac{1}{2}) {\lambda} - \pi/4\bigr)}  + \frac{1}{L-1} \sum_{\ell={L_0/6}}^L O(\ell^{-1})
   =  O(L^{-1}),
\end{align*}
with an implied constant that depends on $\lambda$.
To a sufficient approximation we may therefore neglect $\widehat{q}^{\rm{need}}_L$ compared to $q^{\rm{need}}_\ell$, and so obtain, using \eqref{eq:betadef} and
\eqref{eq:ACdef}
\[
\alpha_k = \frac{\beta_k}{\beta_0},
\]
where
\begin{align*}
\beta_k & \approx \frac{1}{L-1}\sum_{\ell=L_0/6}^{L-k}
q^{\rm{need}}_\ell q^{\rm{need}}_{\ell + k}\\
&\approx \frac{4}{\sqrt{\pi \sin\lambda}}\frac{1}{(L-1)}
\sum_{\ell={L_0/6}}^{L-k}\cos\bigl((\ell+\tfrac{1}{2})\lambda - \pi/4\bigr)\cos\bigl((\ell+k+\tfrac{1}{2})\lambda - \pi/4\bigr)\\
& =\frac{2}{\sqrt{\pi \sin\lambda}}\frac{1}{(L-1)}\sum_{\ell={L_0/6}}^{L-k}
\bigl[\cos(k\lambda)+\cos\bigl((2\ell+k+1)\lambda -\pi/2\bigr)\bigr]\\
& = \frac{2}{(L-1)\sqrt{\pi \sin\lambda}} \bigl[(L-2)
            \cos(k\lambda) + O(1)\bigr]\\
& \approx \frac{2(L-2)}{(L-1)\sqrt{\pi \sin\lambda}} \cos(k\lambda),
\end{align*}
from which it follows that for small $k$ and large $L$,
\begin{equation}\label{eq:akdineed}
\alpha_k = \frac{\beta_k}{\beta_0} \approx \cos(k\lambda),
\end{equation}
where $\cos\lambda=\bp\cdot\bx_0$. Thus in the case of an axially
symmetric needlet the autocorrelation has a distinctive oscillation
as the lag $k$ varies.

But clearly the case $G_{\ell,m}=1$, with its resulting axial symmetry, is not suitable for the present scenario, so we now consider other choices for $G_{\ell,m}$, initially with rotation $\rho$ set equal to the identity matrix $I$. To gain insight into how to choose $G_{\ell,m}$, it is useful to remember that the spherical harmonic $Y_{\ell,0}$ is concentrated at the two poles, see \eqref{eq:Yl0}, whereas, as is well known, $\shY$ for $|m|\approx\ell$ is concentrated near the equator.
With this in mind, we define
\begin{equation}\label{eq:glm}
        G_{\ell,m} := \left\{
        \begin{array}{ll}
        \displaystyle \frac{|m|-\nu\ell}{\ell-\nu\ell}, & |m|\geq \nu\ell,\; \ell\geq0,\\[2mm]
                0, & \mbox{otherwise},
        \end{array}\right.
\end{equation}
for a factor $\nu\in(0,1]$. This choice eliminates all values of $|m|$ smaller than $\nu \ell$, and has the value $1$ for $|m|=\ell$ and in between interpolates linearly. The result is a field that is concentrated near the galactic equator. For our present probes (where we need a field concentrated on the plane $y=0$), we need also a rotation $\rho$ of angle $\pi/2$ about the $x$ axis to change the phase of concentration from the plane $z=0$ to $y=0$.

In summary, our model of the non-random contribution to the field takes
the form \eqref{eq:dineed}, together with \eqref{eq:glm} and the rotation $\rho$ by $\pi/2$ as described in the last paragraph. While we have not proved that our asymmetric needlet has the same approximate scaled autocorrelation as in \eqref{eq:akdineed}, it does appear to be the case.

We need to do Fourier analysis on \eqref{eq:dineed}, so that we construct the corresponding probe coefficients and compute the autocorrelations and AC discrepancies.
To handle the rotation in \eqref{eq:dineed}, we could use the machinery of Wigner D-matrices, but a simpler strategy is available, namely to carry out the computations of probe coefficients and autocorrelation in an unrotated frame, i.e. with $\rho=I$, and then simply rotate the AC discrepancy map.

On setting $\rho=I$ in \eqref{eq:dineed} we then find the Fourier coefficients of $T^{\rm need}$:
\begin{equation}\label{eq:almneed}
        a^{\rm need}_{\ell,m}=\left\{\begin{array}{ll}
                \displaystyle H_{\ell}G_{\ell,m}\conj{Y_{\ell,m}(\bx_0)}, &\hbox{if~} |m|\geq\nu\ell,\; \ell\geq0,\\[1mm]
                0, & \hbox{otherwise}.
        \end{array}\right.
\end{equation}
The probe coefficient $T^{\rm need}_{\ell,\bp}$ for any direction $\bp\in\sph{2}$ is then given by
\begin{equation*}
        T^{\rm need}_{\ell,\bp} = \sqrt{\frac{4\pi}{2\ell+1}}\sum_{|m|\geq\nu\ell}a^{\rm need}_{\ell,m}Y_{\ell,m}(\bp)
        = \sqrt{\frac{4\pi}{2\ell+1}}H_{\ell}\sum_{|m|\geq\nu\ell}G_{\ell,m}\conj{Y_{\ell,m}(\bx_0)}Y_{\ell,m}(\bp).
\end{equation*}

To compute the autocorrelation of the combined field, we first generate a realisation of a random field and then the Fourier coefficients as in Section~4. Meanwhile, the Fourier coefficients of the needlet-like field is given by \eqref{eq:almneed}. By \eqref{eq:T total} and \eqref{eq:Texp}, we then obtain the Fourier coefficients of the combined field $T^{\rm total}$ as
\begin{equation}\label{eq:a total}
        a_{\ell,m}^{\rm total} = a_{\ell,m}^{\rm rand} + \gamma a_{\ell,m}^{\rm need}.
\end{equation}
Using \eqref{eq:probecoeff} and \eqref{eq:a total},
\begin{align*}
        T^{\rm total}_{\ell,\bp} &= T^{\rm rand}_{\ell,\bp} + \gamma T^{\rm need}_{\ell,\bp}
        = \sqrt{\frac{4\pi}{2\ell+1}}\sum_{m=-\ell}^{\ell}\left(a^{\rm rand}_{\ell,m} + \gamma a^{\rm need}_{\ell,m}\right)\shY(\bp),
\end{align*}
and then $q^{\rm total}_{\ell}=\widetilde{T}^{\rm total}_{\ell,\bp}=T^{\rm total}_{\ell,\bp}/\sqrt{\widehat{C}^{\rm total}_{\ell}}$ for $\ell=2,3,\dots,L$, where $L\leq L_0$, and the angular power spectrum is
\begin{equation*}
        C_{\ell}^{\rm total} = \frac{1}{2\ell+1}\sum_{m=-\ell}^{\ell}\left|a^{\rm rand}_{\ell,m} + \gamma a^{\rm need}_{\ell,m}\right|^2.
\end{equation*}
\begin{figure}
\centering
\includegraphics[trim = 0mm 0mm 0mm 0mm, width=0.75\columnwidth]{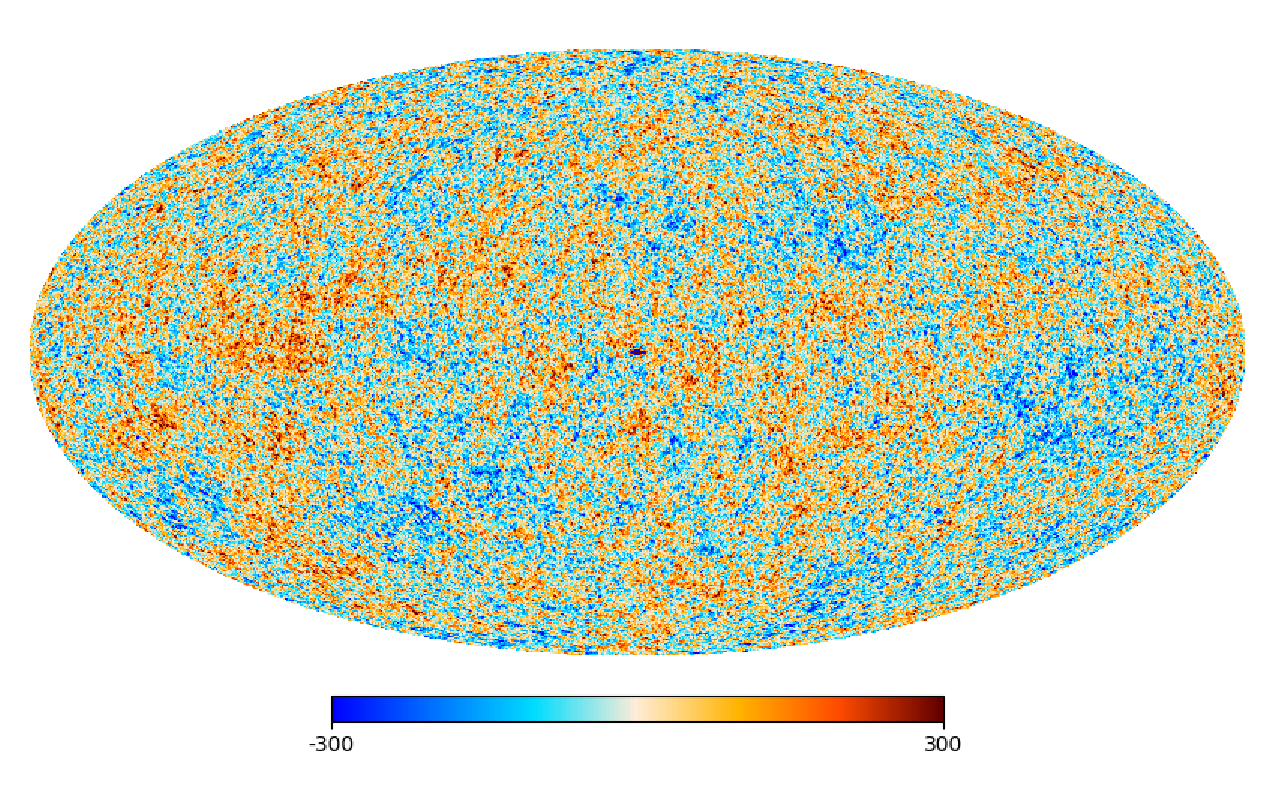}\\[-5mm]
\caption{Model field $T^{\rm rand} + \gamma T^{\rm need}$, $\nu=0.8$, $\gamma=-1.1$,
centre $\bx_0$ at $(l,b)=(0,0)$, $L_0=3,000$ and $H_{\ell}$ as in \eqref{eq:Hell} and \eqref{def:h},
for the map $T^{\rm rand}$ which is instance $1$ of Gaussian random field generated from the best-fit angular power spectrum, $N_{\textrm{Side}} = 2048$}
\label{fig:dineed_instance1}
\vspace{3mm}
\includegraphics[trim = 0mm 0mm 0mm 0mm, width=0.75\columnwidth]{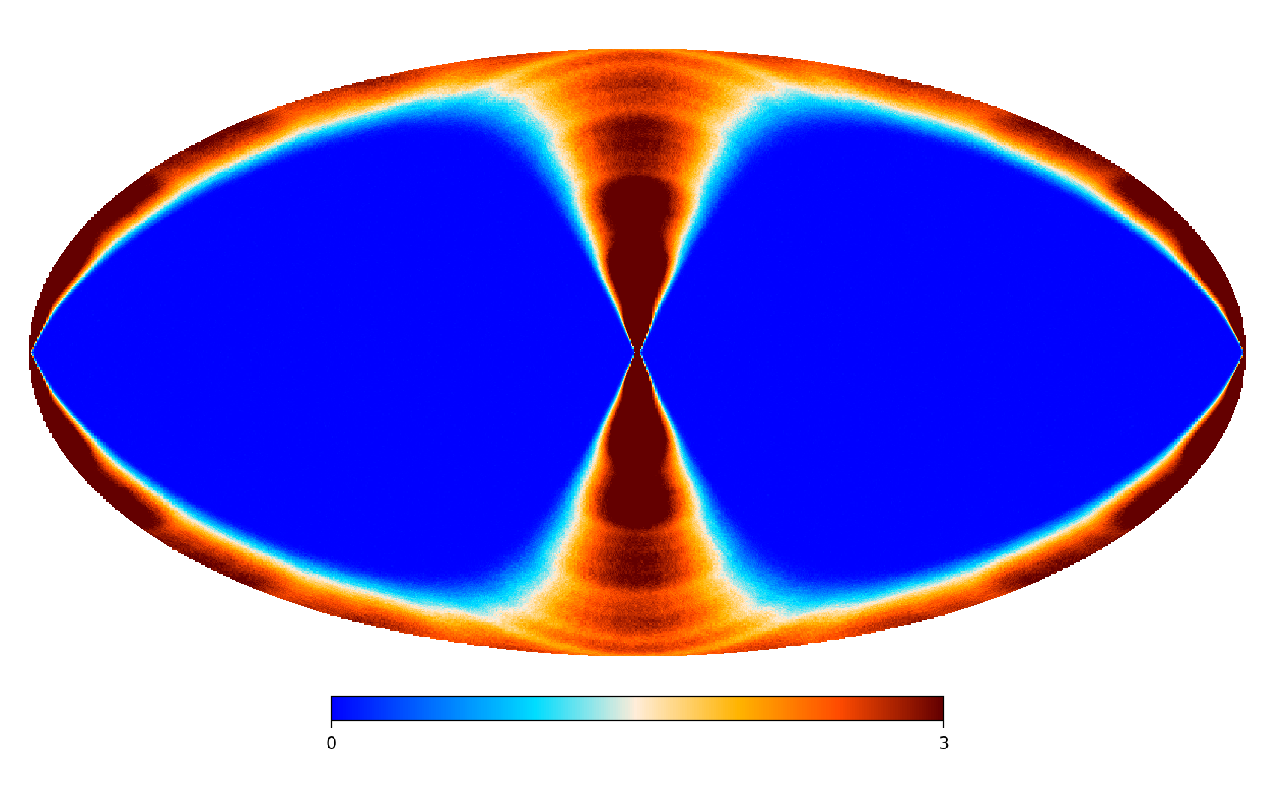}\\[-4mm]
\caption{AC discrepancy map for the additive model field in Figure~\ref{fig:dineed_instance1}, $N_{\textrm{Side}} = 1024$, $k_{\max} = 10$,
${\rm max} = 9.55$ at $(l,b) = (180.00, -0.04)$}
\label{fig:acd.rotdineran}
\end{figure}

\begin{figure}
\vskip 2mm
\centering
\includegraphics[width=0.75\columnwidth]{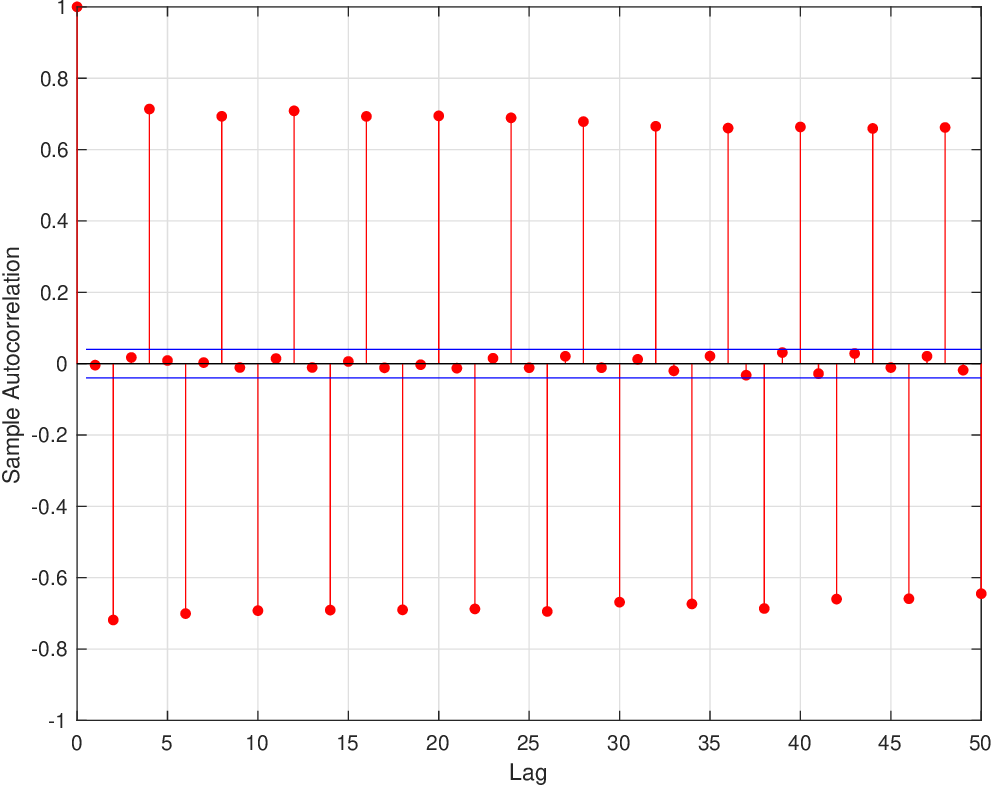}\\[2mm]
\caption{Autocorrelation of $T_{\ell,\bp}/\sqrt{\widehat{C}_{\ell}}$ for \texttt{SEVEM} 2018 with $\bp$ at the North Pole $\bn$}
\label{fig:sev_al0}
\end{figure}
\vskip 2mm
\begin{figure}
\centering
\includegraphics[width=0.75\columnwidth]{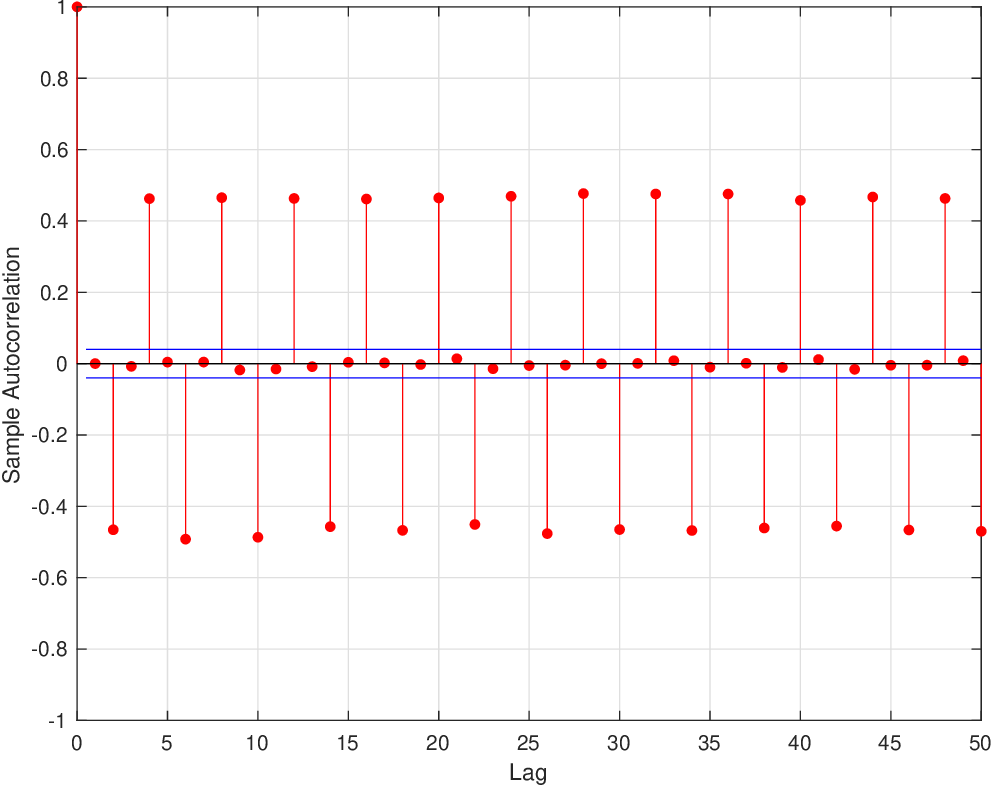}\\[2mm]
\caption{Autocorrelation of probe coefficients of $T^{\rm rand} + \gamma T^{\rm need}$ at the North Pole,
$\nu=0.8$, $\gamma=-1.1$, centre $\bx_0$ at $(l,b)=(0,0)$, $L_0=3,000$}
\label{fig:autocorr_rotdineran_np}
\end{figure}

We now present the numerical results for this model with the following parameters determined by trial and error:
we take $\gamma = -1.1$, $\nu = 0.8$ and $H_\ell$ as in equations~\eqref{eq:Hell} and \eqref{def:h} for all $\ell,m$.
In Figure~\ref{fig:dineed_instance1}, we show the additive model field $T^{\rm total}$ with these parameters, using instance $1$ of the Gaussian random field illustrated in Figure~\ref{fig:grf_instance1}.
Note that the needlet-like structure at the galactic centre is apparently visible.

We give in Figure~\ref{fig:acd.rotdineran} the AC discrepancy map for the model field in Figure~\ref{fig:dineed_instance1}. This is to be compared with Figure~\ref{fig:acd_UIsevem2018} for the case of \texttt{SEVEM} 2018.  We think it fair to claim that the model reproduces to a reasonable extent the principal feature of the \texttt{SEVEM} AC discrepancy, namely the strong band centred on the great circle through the poles and the galactic centre. Note that the scales in the two colour maps are the same. The maximum value of the model AC discrepancy in Figure~\ref{fig:acd.rotdineran} is $9.55$, compared to the maximum AC discrepancy $9.37$ of \texttt{SEVEM} 2018 in Figure~\ref{fig:acd_UIsevem2018}.

In a more detailed comparison of non-inpainted \texttt{SEVEM} 2018 and the model, we examined the autocorrelations on the great circle through the galactic centre and the poles. There turned out to be a strikingly good agreement with the oscillatory behaviour predicted by \eqref{eq:akdineed}. (The prediction is admittedly for an isotropic needlet, but is empirically present also for our directional needlet.)

As an example of the oscillatory behaviour, in Figure~\ref{fig:sev_al0} we
show the observed scaled autocorrelations for non-inpainted \texttt{SEVEM} 2018 at the North Pole. For comparison, we show in
Figure~\ref{fig:autocorr_rotdineran_np} the scaled autocorrelation at $\bp
= \bn$, the {North Pole}, which exhibits the same oscillation. In more detail, for the case $\bp = \bn$ in Figure~\ref{fig:autocorr_rotdineran_np}, for which $\lambda = \pi/2$, the autocorrelations for odd lags $k$ vanish, except for the expected small perturbations. And for even $k$ the values alternate, exactly as expected for the function $\cos(k\pi/2)$.  For small lags $k$ one observes almost the same dependence on $k$ in the case of \texttt{SEVEM} in Figure~\ref{fig:sev_al0}.  The agreement is not as good for the amplitudes of the oscillations, but is perhaps fair, given the simplicity of the model.

Similarly, the non-inpainted \texttt{SEVEM} 2018 autocorrelations showed the oscillatory behaviour everywhere on the great circle through the galactic centre and the poles. Finally, given that the model has a pronounced needlet-like structure at the galactic centre, it is natural to ask if a similar structure is present in the \texttt{SEVEM} 2018 temperature field. The answer is apparently yes: the maximum value of the \texttt{SEVEM} 2018 temperature field has the enormous value of $18161.53$ ({$\mu\mbox{K}$}), occurring close to the galactic centre, at $(l,b)=(0.51,-0.04)$.

\section{Conclusions}\label{sec:conclusion}
In this paper we introduce a `probe' to assess the randomness of
purported isotropic Gaussian random fields, and use the probe
to test the hypothesis that the all-sky CMB temperature anisotropy
maps from the {\it Planck} collaboration are realisations of an isotropic Gaussian field. The probe coefficients for a direction $\bp$ are just coefficients $a_{\ell,0}$ of the field if the $z$-axis is rotated to the direction $\bp$.  Under the assumption that the field is isotropic and Gaussian the probe coefficients for a given direction $\bp$ should be independent Gaussian random variables. Comparing with simulated statistically isotropic Gaussian full-sky maps, we find clear evidence for a departure from this assumption for some of the inpainted {\it Planck} maps, with the \texttt{NILC} 2018 map closest to the Gaussian expectation. The deviations can be made visible by a global computation of the ``AC discrepancy''. Interestingly, we find the excess to lie mostly in the masked region of the maps rather than in the CMB signal-dominated parts, but to be different for each map. It could reflect the variations in the inpainting processes used for the different maps, as well as varying degrees of success in replicating the statistical properties of the unmasked regions and thereby inadvertent violation of statistical isotropy in the full maps.

We also assessed the influence of the maximum multipole $L$ on the AC discrepancies.  Whereas the initial calculations were carried out with $L = 1500$, with the cutoff raised to $L = 2500$, the Commander 2015 AC discrepancy displayed prominent localised anomalies well away from the masked region.

Finally, we applied the probe to the non-inpainted {\it Planck}~2018 maps, obtaining AC discrepancies of a very different kind: instead of the localised ``white noise'' appearance of the AC discrepancies, the AC discrepancies now appear to be continuous and (at least on the evidence of \texttt{SEVEM} 2018), globally dispersed, with a similar AC discrepancy obtained by a model in which in addition to the isotropic random field we add a single narrow needlet-like structure located at the galactic centre. An intuitive explanation of the different behaviour is that the needlet-like structure has a very slowly decaying Fourier spectrum, making the high degree Fourier coefficients of the added deterministic field globally dominant.

We note that even for perfectly Gaussian {\it primordial} perturbations, the observed temperature fluctuations of the CMB are not expected to be Gaussian. It is because the secondary anisotropies generated by non-linear physics introduce a small degree of non-Gaussianity, which would manifest itself as an evenly distributed excess in the AC discrepancy map. It will be interesting to further investigate to what extent these effects contribute to the observed AC discrepancy excess. A further challenge is to extend the analysis to masked maps.

\acknowledgments


Some of the results in this paper have been derived using the \texttt{HEALPix}
\citep{Gorski_etal2005} package.
We acknowledge support from the Australian
Research Council under Discovery Project DP180100506. This material is
based upon work supported by the National Science Foundation under Grant
No.~DMS-1439786 while the authors 3,4,5 were in residence at the Institute
for Computational and Experimental Research in Mathematics in Providence,
RI, during the Point Configurations in Geometry, Physics and Computer
science program. We acknowledge the generous assistance of Domenico
Marinucci, and valuable comments on an earlier draft by Andriy Olenko.

\bibliographystyle{JHEP}
\bibliography{dfp_v1}







\end{document}